\documentclass{aa}  

\usepackage{graphicx,pdflscape,tabularx}
\usepackage{graphicx}
\usepackage{svg}
\usepackage{subcaption}

\usepackage{txfonts}
\usepackage{natbib}
\begin{document}

   \title{MHD Simulations Preliminarily Predict The Habitability and Radio Emission of TRAPPIST-1e}

   \subtitle{}
   
   \author{BoRui Wang\inst{1} , ShengYi Ye\inst{1} \fnmsep\thanks{Just to show the usage
          of the elements in the author field}, J.Varela\inst{2}, \and XinYi Luo\inst{3}
          }

   \institute{Department of Earth and Space Sciences, Southern University of Science and Technology, ShenZhen 518055, P. R. China
         \and
             Institute for Fusion Studies, University of Texas, Austin, TX 78712, USA
        \and 
        State Key Laboratory of Multiphase Flow in Power Engineering, Xi’an Jiaotong University, Xi’an 710049, P. R. China
             }

   \date{submit\today}

% \abstract{}{}{}{}{} 
% 5 {} token are mandatory
 
  \abstract
  % context heading (optional)
  % {} leave it empty if necessary  
   {As the closest Earth-like exoplanet within the habitable zone of the M-dwarf star TRAPPIST-1, TRAPPIST-1e exhibits a magnetic field topology that is dependent on space weather conditions. Variations in these conditions influence its habitability and contribute to its radio emissions.}
  % aims heading (mandatory)
   {Our objective is to analyze the response of different terrestrial magnetosphere structures of TRAPPIST-1e to various space weather conditions, including events analogous to coronal mass ejections (CMEs). We assess its habitability by computing the magnetopause standoff distance and predict the resulting radio emissions using scaling laws. This study provides some priors for future radio observations.}
  % methods heading (mandatory)
   {We perform three-dimensional magnetohydrodynamic (MHD) simulations of the TRAPPIST-1e system using the PLUTO code in spherical coordinates. Assuming TRAPPIST-1e possesses various Earth-like magnetospheres, we analyze the impact of four distinct space weather conditions on its magnetospheric topology. The space weather parameters range from sub-Alfvénic conditions to CMEs.}
  % results heading (mandatory)
   {Our analysis indicates that the predicted habitability and radio emission of TRAPPIST-1e strongly depend on the planet's magnetic field intensity and magnetic axis inclination. Within sub-Alfvénic, super-Alfvénic, and transitional stellar wind regimes, the radio emission intensity positively correlates with both planetary magnetic field strength and axial tilt, while planetary habitability, quantified by the magnetopause standoff distance, shows a positive correlation with magnetic field strength and a negative correlation with magnetic axis tilt. If TRAPPIST-1e possesses a magnetic field strength and axial tilt comparable to Earth's ($\sim 0.32G$ and $\sim 23.5^{\circ}$). its magnetosphere can effectively shield the planetary surface from stellar wind erosion even under severe space weather conditions (e.g., CME events). However, for larger axial tilts ($\sim45^{\circ}$), a stronger planetary magnetic field would be required to ensure adequate protection against intense stellar wind events. Furthermore, our simulations predict radio emission powers up to $\sim10^{20} erg\cdot s^{-1}$ under CME conditions and about $\sim10^{19} erg\cdot s^{-1}$ in the super-Alfvénic regime. This provides a theoretical basis for ground-based direct detection of radio emissions from close-in hot Jupiters, whose strong magnetic fields yield electron cyclotron frequencies exceeding Earth's ionospheric cutoff frequency.}
  % conclusions heading (optional), leave it empty if necessary 
   {}


   \maketitle
%
%-------------------------------------------------------------------

\section{Introduction}
M-dwarf stars represent the most abundant stellar population in the universe, constituting the vast majority of stars within our galaxy and beyond. Their ubiquity and intrinsic characteristics---particularly their low mass, luminosity, and relatively small size---make them compelling targets in the search for potentially habitable exoplanets. Due to their low luminosity, M-dwarfs possess habitable zones located much closer to the star compared to solar-type stars. Moreover, the exceptionally long lifespans of M-dwarfs, often spanning billions to tens of billions of years, provide ample temporal windows conducive to the emergence and evolution of life.

Planets situated within the habitable zones of M-dwarfs may maintain surface temperatures compatible with the presence of liquid water, a fundamental prerequisite for life as we understand it. However, the habitability of these planets is significantly influenced by the host stars' magnetic activity. Young M-dwarfs frequently exhibit intense stellar winds, high levels of X-ray, and ultraviolet (UV) radiation, which can profoundly affect planetary atmospheres. In this context, a planetary magnetic field plays a critical protective role; a sufficiently strong magnetosphere can shield a planet's atmosphere from erosion by stellar winds, thus safeguarding habitability.

Furthermore, given their close orbital proximity, planets around M-dwarfs may exist within the stellar Alfvén surface, a condition conducive to electromagnetic star-planet interactions (SPI). Within this sub-Alfvénic regime, planets can interact magnetically with their host stars, forming Alfvén wings---structures capable of carrying substantial energy fluxes. This process is analogous to the well-studied interaction between Jupiter and its moon Io within our Solar System. Theoretical models such as the 'unipolar inductor' model proposed by \citet{Goldreich}, and the Alfvén wing model developed by \citet{Neubauer} and \citet{Goertz}, have provided foundational frameworks for interpreting such interactions.

Given the substantial differences in stellar properties between M-dwarfs and the Sun---including age, metallicity, magnetic field strength, and rotation rates---the direct application of solar system-based space-weather models to exoplanetary systems remains problematic. If stellar wind conditions (dynamic and magnetic pressures) are intense, planetary habitability becomes heavily dependent on intrinsic planetary magnetic field strength. A robust intrinsic magnetic field can effectively prevent atmospheric stripping by stellar winds, protecting volatile constituents crucial for habitability. Conversely, weak magnetic fields might allow substantial atmospheric loss, diminishing a planet's potential for sustaining life \citep{Jakosky, Airapetian}.

The detection of radio emissions emanating from planetary magnetospheres provides a powerful method to probe these magnetic fields \citep{Hess}. However, current observational capabilities have yet to definitively attribute detected radio signals to either intrinsic planetary magnetospheres or to star-planet interactions. Recent observations, including those of Proxima Centauri \citep{Perez}, Tau Boötis \citep{Turner}, and YZ Ceti \citep{Pineda, Trigilio}, have provided intriguing yet inconclusive evidence.

In particular, the TRAPPIST-1 system offers a significant opportunity to explore these phenomena due to its unique configuration of multiple Earth-sized planets orbiting within or near the habitable zone of a very-low-mass M8 star \citep{Gillon2016, Gillon2017}. TRAPPIST-1e, specifically located within its star’s habitable zone, serves as an ideal candidate for investigating habitability conditions and radio emission scenarios driven by SPI. Observations of TRAPPIST-1 have suggested active stellar coronae and chromospheres \citep{Bourrier2017, Wheatley2017}. Variations in stellar wind conditions and Alfvén surface locations predicted by different models \citep{Garraffo2017, Dong2018} further emphasize the complexity of assessing habitability in this system.

Recent studies indicate significant variability in expected radio emissions depending on planetary orbital positions relative to the stellar Alfvén surface, highlighting potential SPI-driven processes in the TRAPPIST-1 system \citep{Reville2024, Perez2021}. TRAPPIST-1e, owing to its proximity and orbital characteristics, is particularly suited for detailed investigation of these interactions. Systematic studies of TRAPPIST-1e's magnetic environment and associated radio emissions could provide crucial insights into planetary magnetosphere configurations, stellar-planetary electromagnetic interactions, and their implications for habitability.

In summary, detailed investigations of TRAPPIST-1e and similar M-dwarf planetary systems advance our understanding of exoplanetary environments, star-planet electromagnetic interactions, and habitability conditions. Future radio observations of systems like TRAPPIST-1 hold promise for significantly advancing our capabilities to characterize exoplanetary magnetospheres and potentially detect signatures indicative of habitable conditions beyond the Solar System.

In this paper, we investigate the magnetic response of the TRAPPIST-1e system under various terrestrial-like magnetic field strengths and different space weather conditions, including the impact of stellar winds on planetary habitability and associated radio emission. Our methodology closely follows previous studies dedicated to analyzing how space weather affects exoplanetary habitability and planetary magnetospheric radio emissions \citep{Varela2022a,Varela2022b}, which have also been applied to the Proxima b planetary system \citep{Luis2024}. The structure of the paper is as follows: In Section \ref{section2}, we introduce the single-fluid MHD numerical model employed in our simulations. In Section \ref{section3}, we analyze the effects of various space weather conditions on the habitability of TRAPPIST-1e. Section \ref{section4} presents calculations of predicted radio emissions under different scenarios. Finally, Section \ref{section5} summarizes our findings and provides a discussion.

%--------------------------------------------------------------------
\section{Numerical Model of PLUTO}
\label{section2}

We employed the ideal magnetohydrodynamics (MHD) module of the open-source PLUTO code \citep{Mignone2007} in spherical coordinates to model the time evolution of a single-fluid, fully ionized polytropic plasma under the non-resistive and inviscid approximation. The system of equations solved includes mass, momentum, magnetic induction, and total energy conservation, expressed in conservative form as:

\begin{align}
\frac{\partial \rho}{\partial t} + \nabla \cdot (\rho \mathbf{v}) &= 0, \\
\frac{\partial \mathbf{m}}{\partial t} + \nabla \cdot \left[ \mathbf{m} \mathbf{v} - \frac{\mathbf{B}\mathbf{B}}{\mu_0} + \mathbf{I} \left( p + \frac{B^2}{2\mu_0} \right) \right]^T &= 0, \\
\frac{\partial \mathbf{B}}{\partial t} + \nabla \times \mathbf{E} &= 0, \\
\frac{\partial E_t}{\partial t} + \nabla \cdot \left[ \left( \frac{1}{2} \rho \mathbf{v}^2 + \rho e + p \right) \mathbf{v} + \frac{\mathbf{E} \times \mathbf{B}}{\mu_0} \right] &= 0,
\end{align}

where $\rho$ is the mass density, $\mathbf{v}$the velocity field, $\mathbf{m}=\rho v$ the momentum density,  $\mathbf{B}$ the magnetic field,  $\mathbf{E}=-(\mathbf{v}\times\mathbf{B})$ the electric field, $p$ the thermal pressure and $E_{t}=\rho e+\frac{1}{2}\rho \mathbf{v}^2+\frac{\mathbf{B}^2}{2\mu_{0}}$ the total energy density. The internal energy $\rho e$ is closed via the ideal gas equation of state:
\begin{equation}
\rho e = \frac{p}{\gamma - 1},
\end{equation}
where $\gamma=5/3$ is the adiabatic index.

For a fully ionized proton-electron plasma, we adopt a mean molecular weight $\mu=1/2$ , so the number density is:
\begin{equation}
n = \frac{\rho}{\mu m_p},
\end{equation}
and the pressure is:
\begin{equation}
p = n k_B T,
\end{equation}
where $k_B$ is Boltzmann's constant and $T$ the temperature. The sound speed is defined as:
\begin{equation}
c_s = \sqrt{\frac{\gamma p_t}{\rho}},
\end{equation}
where $p_t$ is the total proton and electron pressure.

The equations were integrated using a Harten–Lax–van Leer (HLL) approximate Riemann solver with a diffusive minmod limiter. The magnetic field was initialized to be divergence-free and maintained as such throughout the simulations via a mixed hyperbolic/parabolic divergence-cleaning technique \citep{Dedner2002}.
\subsection{Basic setting}

The simulation domain is defined as a three-dimensional spherical grid composed of 128 radial cells, 48 polar cells uniformly distributed in $\theta\in[0,\pi]$, and 96 azimuthal cells covering $\phi\in[0,2\pi]$. The computational domain consists of a concentric shell centered on TRAPPIST-1e, with the radial range extending from the inner boundary $R_{in}=2.0R_e$  to the outer boundary $R_{out}=30R_e$ , where $R_e=0.918R_{\oplus}$ is the radius of TRAPPIST-1e \citep{Gillon2016}. The radial grid spacing is uniform, and all simulations assume a characteristic stellar wind speed $V=10^7cm/s$.

An upper ionosphere model is implemented between $R_{in}$ and $R=2.2R_{e}$, prescribing plasma inflow driven by field-aligned currents. This region provides a critical boundary layer where electric fields influence plasma motion. To prevent artificial numerical reflections, the outer boundary is divided into an upstream region with fixed stellar wind parameters and a downstream region with zero radial derivative conditions $\partial/\partial r=0$ for all variables.

Initial conditions are set with a cutoff to the interplanetary magnetic field (IMF) at $R_{cut}=6R_{e}$, approximating the location where the planetary magnetic pressure dominates over the stellar wind. A density cavity is introduced in the dayside magnetosphere according to:
\begin{equation}
x < R_{\mathrm{cut}} - \frac{y^2+z^2}{R_{\mathrm{cut}}*\sqrt{R_{\mathrm{cut}}}},
\end{equation}
with (x,y,z) are the Cartesian coordinates,where the plasma velocity is zero. The density profile is adjusted to maintain a constant Alfv\'en speed:
\begin{equation}
v_A = \frac{B}{\sqrt{\mu_0 \rho}},
\end{equation}
with $\rho=n_{sw}m_{p}$, where $n_{sw}$ is the stellar wind particle density. To ensure computational feasibility, $v_A$ is fixed at $\sim10^4km/s$, setting the time step for each simulation.

The planetary magnetic field is modeled as a dipole, rotated by 90$^\circ$ in the YZ-plane relative to the grid to avoid numerical singularities at the poles. The simulation frame is defined such that the magnetic axis of TRAPPIST-1e aligns with the z-axis, the star-planet line lies in the XZ-plane (with the star at $x<0$), and the y-axis completes the right-handed system. The model assumes that the planet's magnetic and rotation axes are aligned. Obliquity effects are introduced by adjusting the direction of the IMF and stellar wind vectors.

This setup reproduces major magnetospheric structures, including the bow shock, magnetopause, and magnetosheath, as demonstrated in prior studies \citep{Varela2015,Varela2016a,Varela2016b,Varela2018}. Magnetic reconnection between the planetary and interplanetary fields is assumed to occur instantaneously, due to the enhanced numerical magnetic diffusivity, resulting in an erosion of the planetary magnetic field. Although plasma depletion layers are not resolved due to grid limitations, the simulation captures the essential global features of the magnetosphere.

Simulations are evolved until a steady state is reached, typically after $\tau=L/V\approx15$ code units, equivalent to $t\approx16$ minutes of physical time. This timescale corresponds to the Alfv\'en crossing time over the domain, sufficient to resolve the steady response of the magnetosphere to constant space weather conditions. The transient response to evolving space weather, planetary rotation, and orbital motion are not included in this study and will be addressed in future work.
\subsection{Upper Ionosphere Model}

The upper ionospheric domain in our simulations is defined between radial distances of $R=2.0R_e$ and $R=2.2R_e$ , following the approach of \citet{Buchner2003}. This region is critical for prescribing plasma inflow from field-aligned currents (FACs) into the simulation domain. Below this boundary, the magnetic field intensity becomes too large, resulting in prohibitively small simulation time steps. Moreover, since the single-fluid MHD approximation fails to capture kinetic processes prominent in the inner ionosphere and plasmasphere, these inner regions are excluded from the simulation domain.

To compute the FACs, we begin by evaluating the total current:
\begin{equation}
\mathbf{J} = \frac{1}{\mu_0} \nabla \times \mathbf{B},
\end{equation}
then extract the perpendicular current component:
\begin{equation}
\mathbf{J}_\perp =\mathbf{J}- \left( \mathbf{J} \cdot \mathbf{B} \right) \frac{\mathbf{B}}{|\mathbf{B}|^2},
\end{equation}
leading to the field-aligned current:
\begin{equation}
\mathbf{J}_\mathrm{FAC} = \mathbf{J} - \mathbf{J}_\perp.
\end{equation}

The electric field in the upper ionosphere is derived using an empirical Pedersen conductivity $\sigma$, defined as:
\begin{equation}
\sigma = \frac{40 E_0 \sqrt{F_E}}{16 + E_0^2},
\end{equation}
where $E_0=k_BT_e$ is the mean electron energy, and $F_E = n_e \sqrt{E_0 / (2\pi m_e)}$ is the energy flux. The electric field is then:
\begin{equation}
\mathbf{E} = \frac{\mathbf{J}_\mathrm{FAC}}{\sigma}.
\end{equation}
Plasma drift velocity in this region is calculated from the standard  relation:
\begin{equation}
\mathbf{u} = \frac{\mathbf{E} \times \mathbf{B}}{|\mathbf{B}|^2}.
\end{equation}

To control the simulation time step, we impose a fixed Alfv\'en speed $v_A$ by prescribing the density profile as:
\begin{equation}
\rho = \frac{|\mathbf{B}|^2}{\mu_0 v_A^2},
\end{equation}
where $v_A$ is set $\sim10^4km/s$ depending on the IMF strength and stellar wind density. This ensures numerical stability across different space weather configurations.

The pressure profile is initialized with a smooth transition to the simulation domain and defined in relation to the sound speed:
\begin{equation}
p = n \frac{1}{\gamma} \left[ (c_p - c_{sw}) \frac{r^3 - R_s^3}{R_{\mathrm{un}}^3 - R_s^3} + c_{sw} \right]^2,
\end{equation}
where $\gamma=5/3$,$c_p = \sqrt{\gamma k_B T_p / m_p}$, $ c_{sw} = \sqrt{\gamma k_B T_{sw} / m_p}$, and $T_p$, $T_{sw}$ are the plasma and stellar wind temperatures, respectively.

During the early stages of the simulation, the pressure and density gradients generate plasma outflows from the upper ionosphere into the domain until a quasi-steady state is established. This process is driven by the balance between stellar wind injection and ionospheric outflow. 

We benchmarked the model against the global magnetosphere configuration described by \citet{Samsonov2016}. Using similar parameters (e.g. $n=5cm^{-3},V_x=-400km/s, T = 2 \times 10^5\mathrm{K}, B_y = -B_x = 3.5\mathrm{nT} $), the predicted magnetopause locations $R_x = 10.7R_e, R_y = 16.8R_e, R_z = 14.9R_e$  show good agreement with their results. The computed electric fields near the bow shock and inside the magnetosheath are consistent with Cluster observations \citep{DeKeyser2005}, except in cases with extreme IMF strength.

For a Carrington-like event \citep{Ridley2006} with $n = 750\mathrm{cm}^{-3}, V_x = -1600\mathrm{km/s}, T = 3.5 \times 10^7\mathrm{K}, |\mathbf{B}| = 200\mathrm{nT}$, our model predicts $R/ R_e \sim 1.22$ by extrapolation, which matches within the expected limits.

Finally, we find that the electric field structure in the upper ionosphere remains relatively unchanged due to the fixed density. Both the magnitude and orientation of FACs are within observational ranges (nA/m$^2$ to $\mu$A/m$^2$), in line with previous studies \citep{Weimer2001,Waters2001,Ritter2013,Bunescu2019,Zhang2020}.
%---------------------------------------------------------------------
\section{The Habitability of TRAPPIST-1e}
\label{section3}
\begin{figure}
    \centering
    \includegraphics[width=\linewidth]{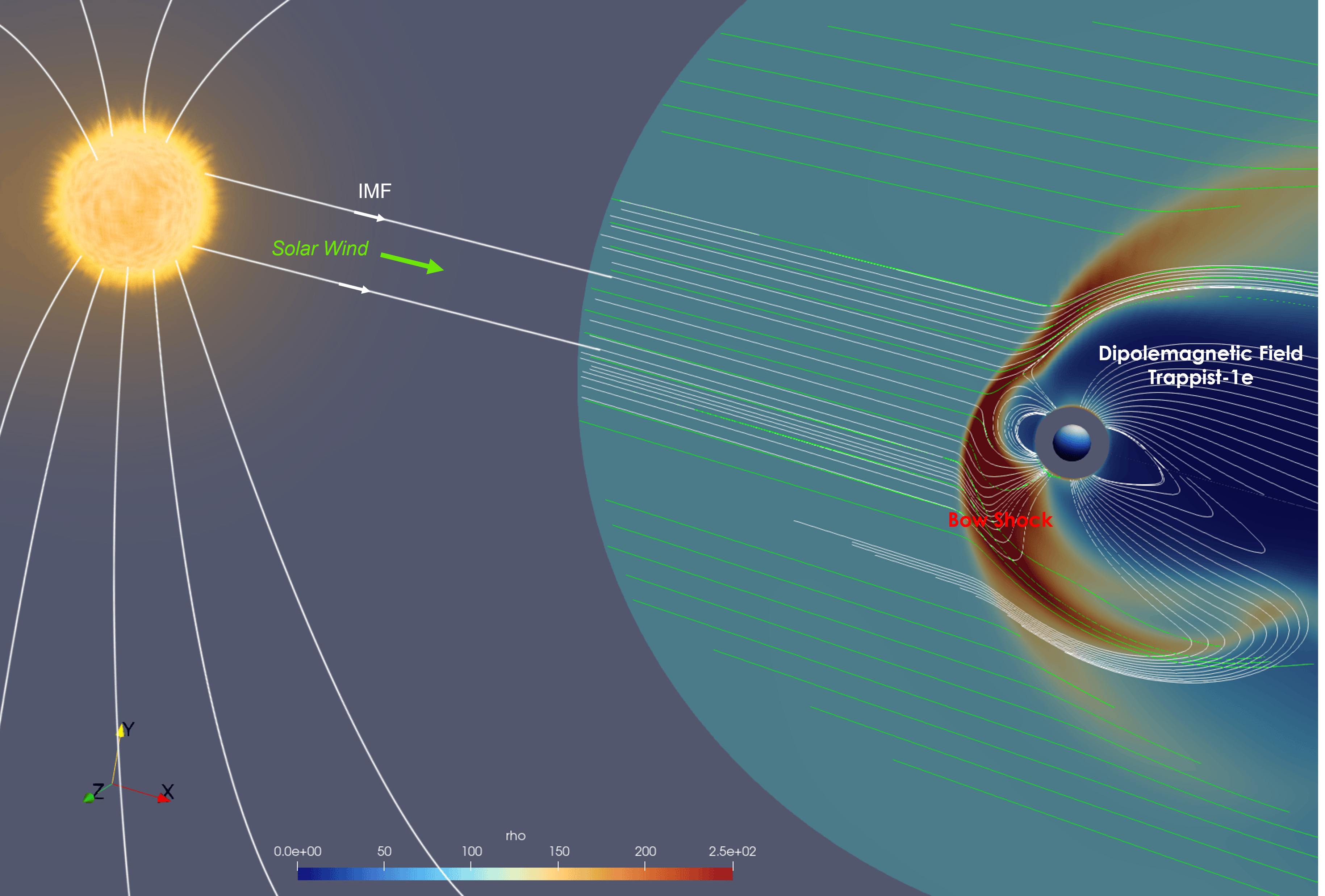}
    \caption{Sketch of the magnetospheric interaction in the TRAPPIST-1--TRAPPIST-1e system. The stellar wind velocity and IMF streamlines are drawn in green and white, respectively. The M dwarf star is shown as a yellow sphere, and the planet TRAPPIST-1e as a blue shaded circle. The planetary magnetic field lines (white) exhibit compression on the dayside and extension into a magnetotail. A bow shock structure forms upstream of the planet under a super-Alfv\'enic regime. The color scale represents normalized plasma density.}
    \label{Figure1}
\end{figure}

Figure~\ref{Figure1} presents a 3D schematic of the TRAPPIST-1e system, including the M dwarf TRAPPIST-1, the stellar wind (green streamlines), interplanetary magnetic field (white streamlines), the planet TRAPPIST-1e (right), and its dipolar magnetic field lines (also white). The stellar wind is assumed to be radial, and the planetary magnetic axis is aligned such that the rotational axis of TRAPPIST-1e is perpendicular to the ecliptic (tilt = $0^{\circ}$). On the dayside, interaction between the stellar wind and the planetary magnetic field leads to the formation of a bow shock (BS), marked by plasma pile-up and density enhancement in red. The stellar wind compresses the planetary field on the dayside and stretches it into a magnetotail on the nightside. Reconnection between the IMF and the planetary magnetic field occurs near the magnetopause, driving localized magnetic erosion and flux transport.

\subsection{Parameter settings in simulation}

We explore a range of space weather scenarios classified into calm and extreme regimes. The calm regime includes three distinct subcases: (1) a sub-Alfv\'enic scenario with a stellar surface magnetic field of 1200 G and low wind dynamic pressure ($M_A < 1$), (2) a transition scenario with reduced stellar magnetic field (600 G), and (3) a super-Alfv\'enic case where dynamic pressure dominates over magnetic pressure, forming a well-defined bow shock. The extreme space weather case simulates a CME-like event with enhanced wind velocity, density, and magnetic field, scaled from the super-Alfv\'enic case by factors of 2.5, 5, and 10, respectively, following \citep{Luis2024}.

 Table \ref{table1} summarises the simulation parameters for all scenarios. The transition scenario was recalculated according to the parameters of the sub-Alfv\'en region by reducing the magnetic field from 1200 G to 600 G. All other space weather condition parameters were adjusted to the orbital environment of TRAPPIST-1e according to the parameter settings for Proxima b planets in \citep{Luis2024} (assuming that the TRAPPIST-1 stars have the same mass loss rate as Proxima Cen). In all cases, the IMF is assumed to be radial.

\begin{table*}[ht]
\caption{Input parameters used for the four simulated TRAPPIST-1e space weather scenarios.}
\label{table1}
\centering
\begin{tabularx}{\textwidth}{lXXXX}
\hline\hline
Parameter & Sub-Alfvénic & Transition & Super-Alfvénic & CME-like \\
\hline
Scenario type & Calm & Calm & Calm & Extreme \\
$n_{\mathrm{sw}}$ (cm$^{-3}$) & 150 & 120 & 150 & 745 \\
$V_{\mathrm{sw}}$ ($10^7$ cm/s) & 4.6 & 3.65 & 9.2 & 23 \\
$B_{\mathrm{IMF}}$ (nT) & 530 & 365 & 265 & 2650 \\
$B_{\mathrm{TRAPPIST\text{-}1}}$ (G) & 1200 & 600 & 600 & 600 \\
$B_{\mathrm{TRAPPIST\text{-}1e}}$ (G) & 0.32--1.28 & 0.32--1.28 & 0.32--1.28 & 0.32--1.28 \\
Tilt ($^\circ$) & 0--45 & 0--45 & 0--45 & 0--45 \\
\hline
\end{tabularx}
\end{table*}

The planetary magnetic field is assumed to be dipolar with strengths of 0.32, 0.64, and 1.28 G (Earth-like to 4$\times$ Earth), and the stellar wind temperature is fixed at $T_{SW} = 2.6 \times 10^5$~K. Simulations explore various tilt angles from $0^{\circ}$ to $45^{\circ}$ to evaluate the impact of magnetic obliquity on the magnetospheric response and radio emission.

\begin{figure}[ht]
\centering
\includegraphics[width=0.97\linewidth]{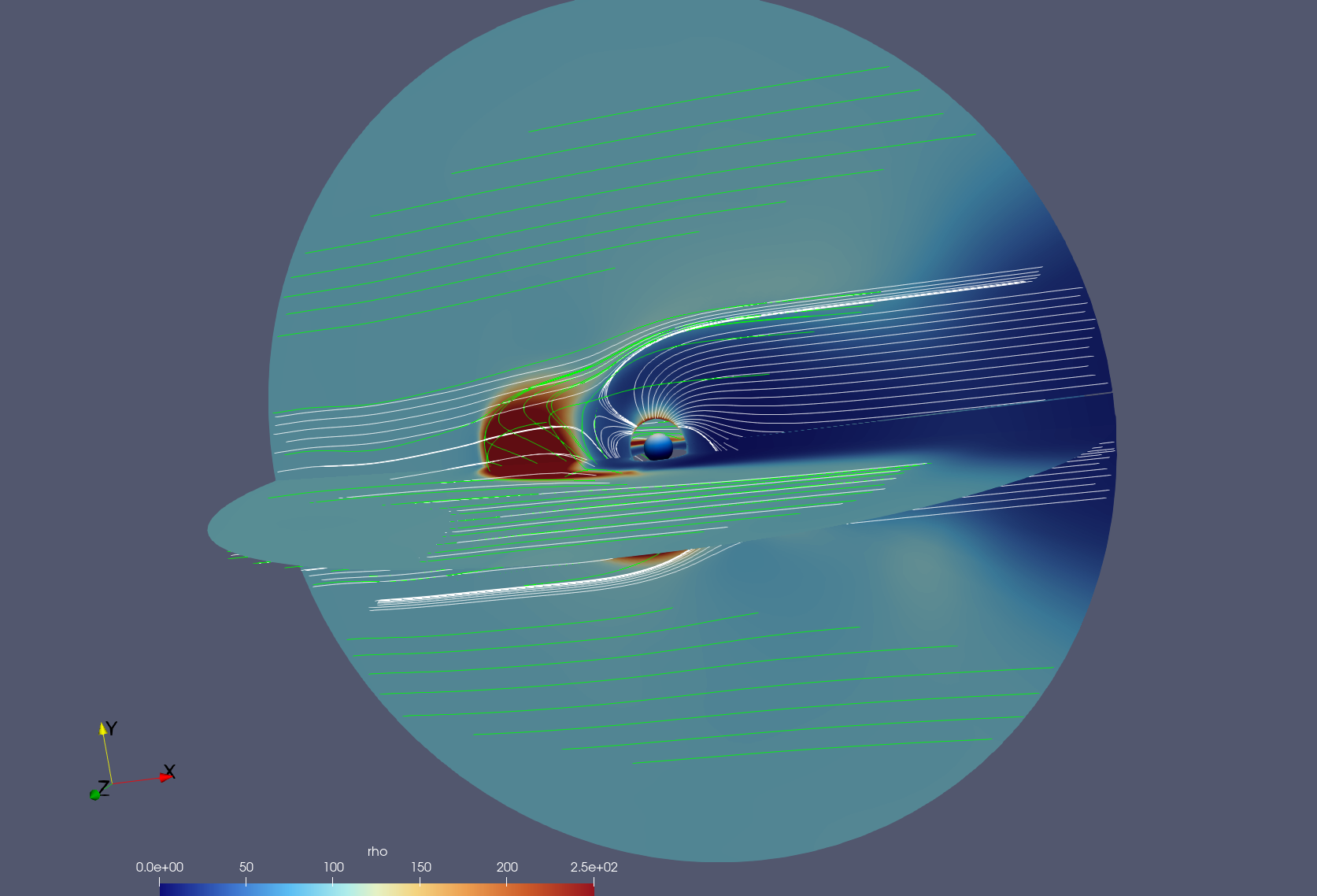}\
\includegraphics[width=0.97\linewidth]{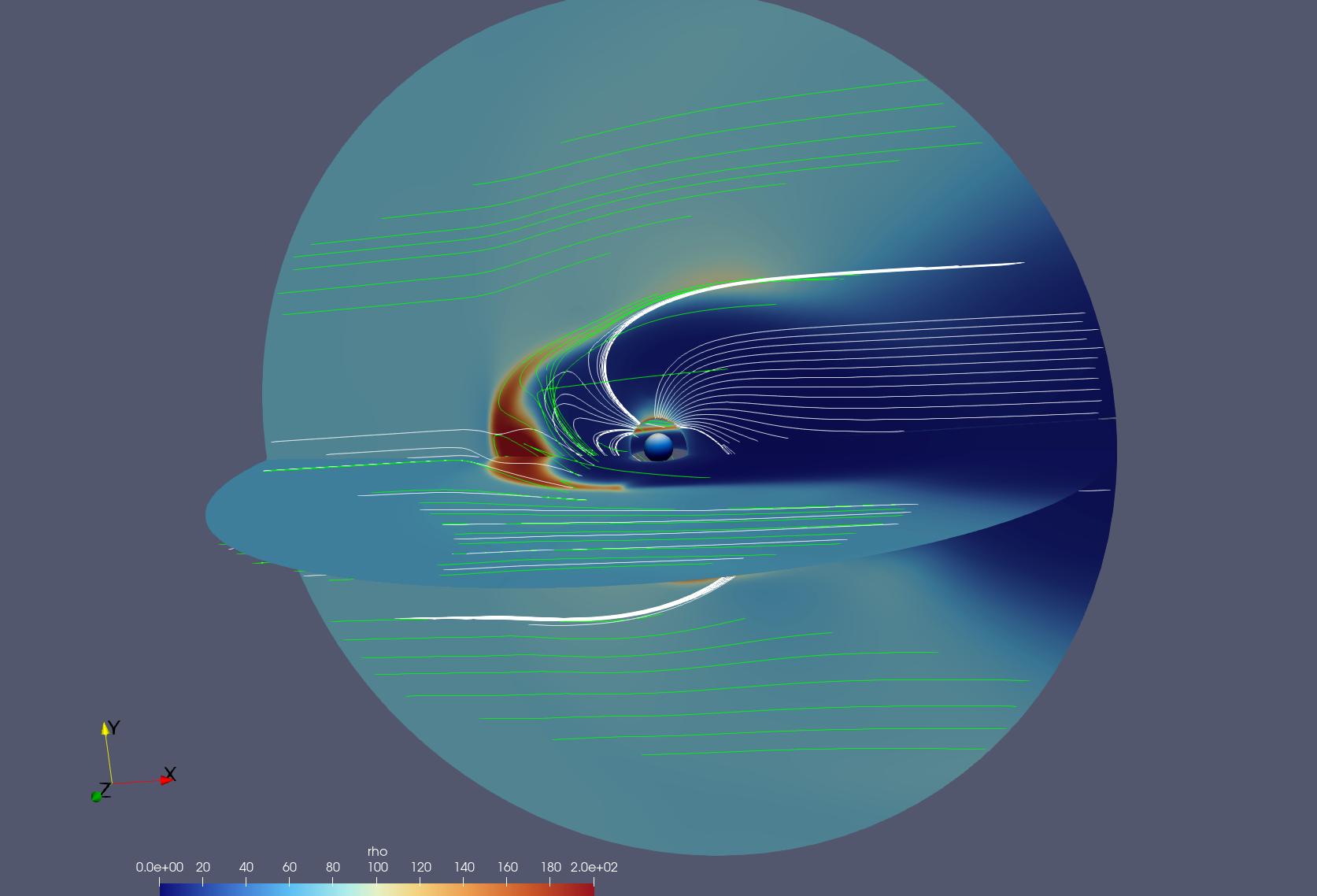}\
\includegraphics[width=0.97\linewidth]{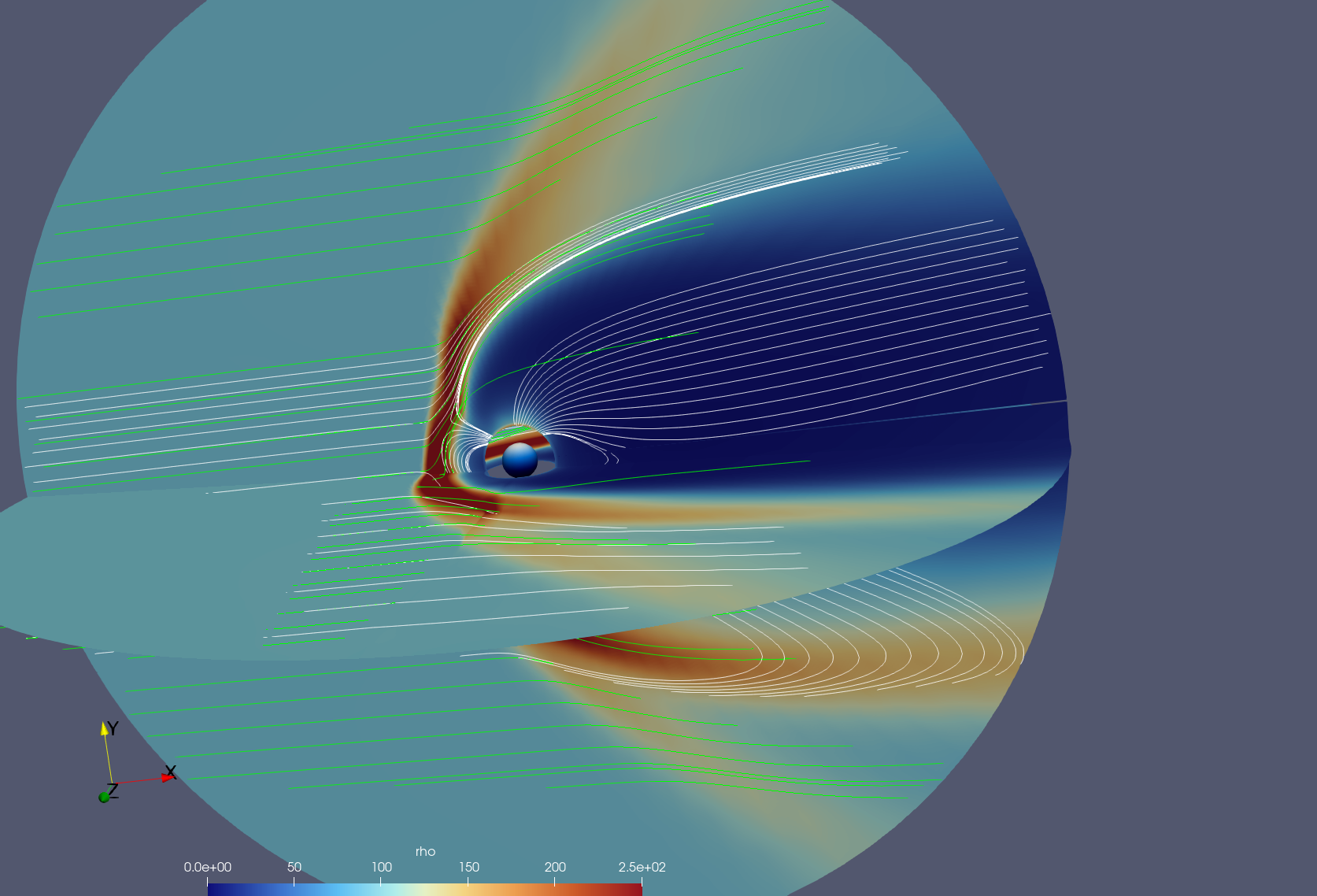}
\caption{Magnetospheric structures under (upper) sub-Alfv\'enic, (middle) transition, and (bottom) super-Alfv\'enic conditions.  The colour scale represents the density distribution of the solar wind. In the sub-Alfv\'enic regime ($M_A < 1$), the stellar wind does not form a bow shock and the radio emission originates solely from magnetospheric reconnection. In the transition case, a weak and asymmetric shock begins to form. In the super-Alfv\'enic case ($M_A > 1$), a clear bow shock develops, and the radio emission is enhanced by both the shock and the reconnection region.}
\label{fig:3panels}
\end{figure}

 Figure \ref{fig:3panels} shows simulations of TRAPPIST-1e under three calm space weather conditions: the sub-Alfv\'en case, the transition case, and the super-Alfv\'en case.The sub-Alfv\'en case (top left) in a 1200 G stellar magnetic field shows no bow shocks; the interaction is magnetically dominated. The transition case (top right) maintains similar wind parameters, but the magnetic field decreases (600 G), producing a slight asymmetry and a weak bow shock structure near the planetary nose. The super-Alfv\'en case (lower panel) leads to a well-defined bow shock and a significant compression of the heliospheric magnetosphere.

\begin{figure}[ht]
\centering
\includegraphics[width=0.48\linewidth]{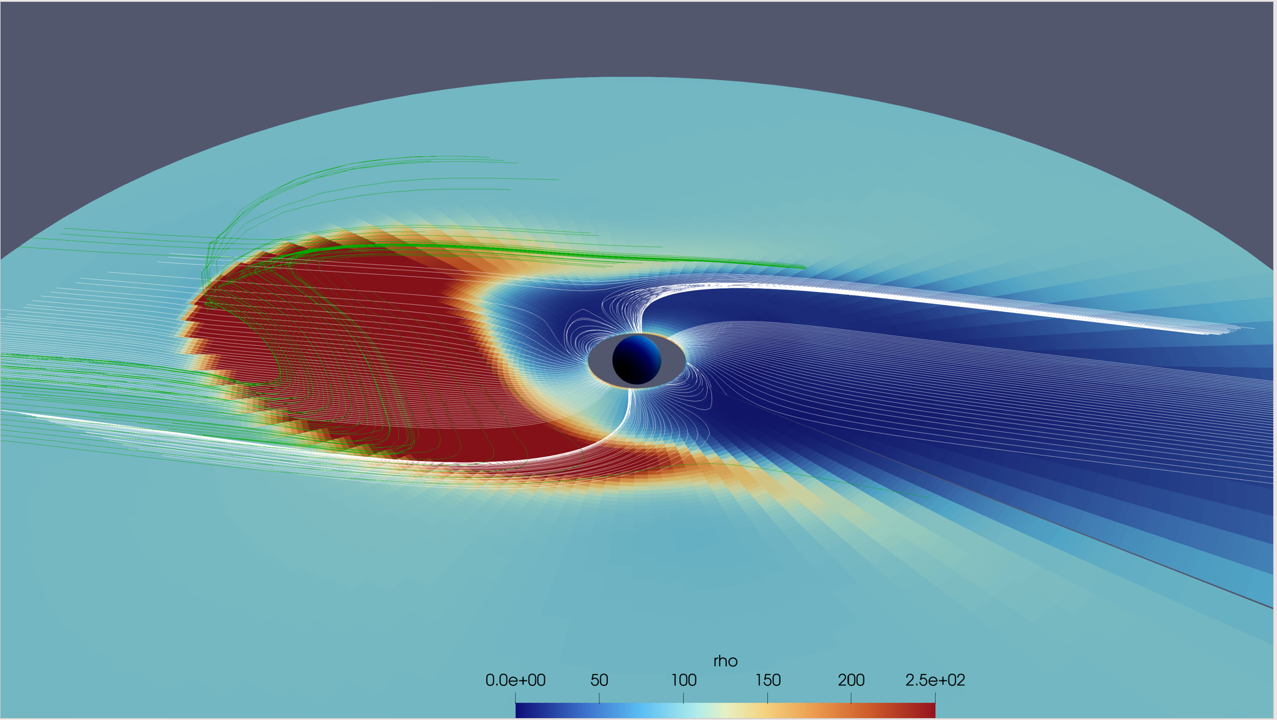}
\includegraphics[width=0.48\linewidth]{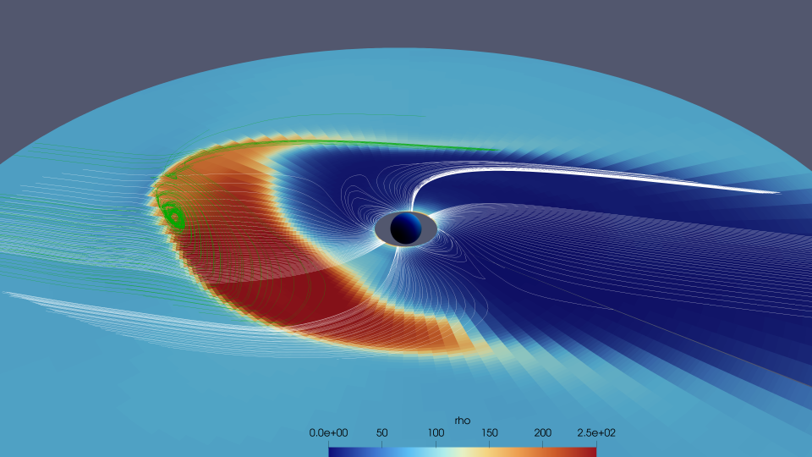}\
\includegraphics[width=0.48\linewidth]{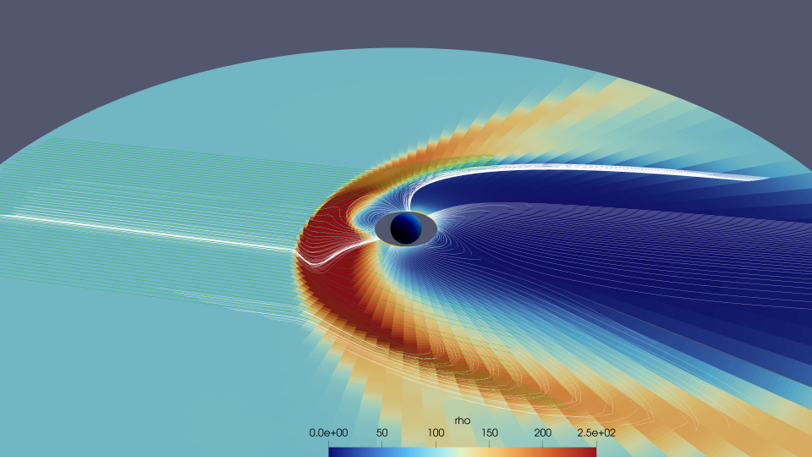}
\includegraphics[width=0.48\linewidth]{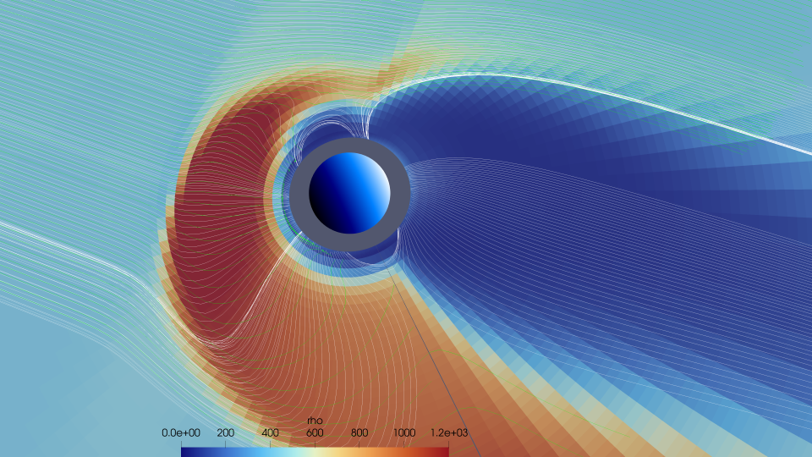}
\caption{Equatorial cuts of the TRAPPIST-1e magnetosphere for four space weather scenarios. Color indicates normalized plasma density; green lines trace stellar wind streamlines, white lines denote planetary and interplanetary magnetic fields. Top left: sub-Alfv\'enic case, no bow shock; top right: transition, weak shock forms; bottom left: super-Alfv\'enic, strong bow shock; bottom right: CME-like event, highly compressed and asymmetric magnetosphere.}
\label{fig:4panels}
\end{figure}

 Figure \ref{fig:4panels} shows zoomed-in maps of simulations on the equatorial plane for four space weather conditions: sub-Alfv\'enic (top left), transition (top right), super-Alfv\'enic (bottom left), and CME-like (bottom right). The color scale represents normalized plasma density. Sub-Alfv'enic conditions show magnetically dominated interaction without a magnetosheath. In the transition regime, a tenuous magnetosheath begins to appear as the magnetic control weakens. Under super-Alfv\'enic and CME-like forcing, dense magnetosheaths and strongly compressed magnetospheres are evident.

\subsection{Habitability Judgement - Magnetopause Standoff Distance}
\label{section3.2}
In this study, we use the magnetopause standoff distance $R_{msd}$ as a diagnostic criterion to evaluate the potential habitability of TRAPPIST-1e, treating it as a function of both the planetary magnetic field strength and the tilt angle of the magnetic axis. In general, if $R_{msd}\leq R_e$, it implies that the planetary magnetosphere is unable to effectively shield the planetary surface from the impinging stellar wind. In this case, solar wind particles may directly precipitate onto the surface, significantly degrading the habitability conditions due to atmospheric erosion.

For simulations of extreme space weather—particularly CME-like scenarios—we set the minimum inner boundary to $R_{in}=1.5R_e$, below which numerical instabilities may occur. Therefore, in such extreme cases, we consider TRAPPIST-1e unprotected if $R_{msd}\leq1.5R_e$, as the magnetosphere fails to prevent direct stellar wind impact.

The value of $R_{msd}$ is determined from the simulation outputs by measuring the last closed magnetic field line in the substellar direction, as adopted in previous works \citep{Varela2022b, Luis2024}, following the same methodology used in the analysis of Ganymede’s magnetosphere \citep{Kivelson2004}. This approach is simulation-based and does not rely on theoretical expressions. The theoretical standoff distance can be derived by equating the total pressure of the stellar wind with the pressure provided by the planetary magnetosphere. The stellar wind pressure includes the thermal pressure $P_{sw,th}=m_pn_{sw}c_{sw}^2/\gamma$,  magnetic pressure $P_{sw,m}=B_{sw}^2/2\mu_{0}$, and dynamic pressure $P_{d}=m_{p}n_{sw}v_{sw}^2/2$. The planetary side includes its thermal pressure $P_{th,e}=m_pn_{e}v_{th,e}^2/2$, and magnetic pressure expressed as $P_{m,e}=\alpha\mu_0 M_{e}^2/8\pi^2r^6$.where $M_e$ is the planetary magnetic moment, $\alpha$ is dipole compression coefficient ($\alpha \approx2$ \citep{Gombosi1994}), and $\mu_0$ is the magnetic permeability.

Equating the total pressures, the theoretical expression for the normalized magnetopause standoff distance becomes:
\begin{equation}
\frac{R_{\mathrm{msd}}}{R_p} = \left[ \frac{ \alpha M_p^2/\pi }{ m_p n_{\mathrm{sw}} v_{\mathrm{sw}}^2 + \frac{B_{\mathrm{IMF}}^2}{4\pi} + \frac{2 m_p n_{\mathrm{sw}} c_{\mathrm{sw}}^2}{\gamma} - m_p n_e v_{\mathrm{th,e}}^2 } \right]^{1/6}, \label{eq:Rmsd}
\end{equation}

It is important to emphasize that Eq.~(\ref{eq:Rmsd}) provides only a rough theoretical estimate of the  magnetopause distance \( R_{\mathrm{msd}} \), and may significantly deviate from the values obtained through global simulations or inferred from observations. Several key limitations must be considered when applying this expression. First, the derivation neglects the magnetic topology of the system. All terms in Eq.~(\ref{eq:Rmsd}) are treated as scalar quantities, which oversimplifies the inherently vectorial and three-dimensional nature of the stellar wind–magnetosphere interaction. In particular, the orientation of the interplanetary magnetic field (IMF) relative to the planetary dipole is not accounted for. As a result, the formula assumes a purely compressed dipole geometry, which does not represent realistic reconnection-driven magnetospheric configurations. Second, Eq.~(\ref{eq:Rmsd}) does not include the effects of IMF–planetary field magnetic reconnection, which plays a critical role in controlling magnetospheric erosion and standoff distance.  Third, while the theoretical expression yields a unique value for any given set of parameters \( n_{\mathrm{sw}} \), \( v_{\mathrm{sw}} \), \( B_{\mathrm{IMF}} \), and \( B_p \), our three-dimensional MHD simulations reveal a range of magnetopause standoff distances depending on the tilt angle and the specific orientation of the IMF. This inherent variability reflects the complexity of magnetospheric dynamics and underlines the limitations of applying Eq.~(\ref{eq:Rmsd}) as a general diagnostic. In extreme space weather cases or under strong reconnection conditions, the discrepancy between the theoretical and simulated values may be substantial. Therefore, while Eq.~(\ref{eq:Rmsd}) can serve as an approximation, accurate determination of $R_{\mathrm{msd}}$ requires 3D MHD modeling that includes both magnetic topology and reconnection, as presented in this work.

\begin{figure}
    \centering
    \includegraphics[width=\linewidth]{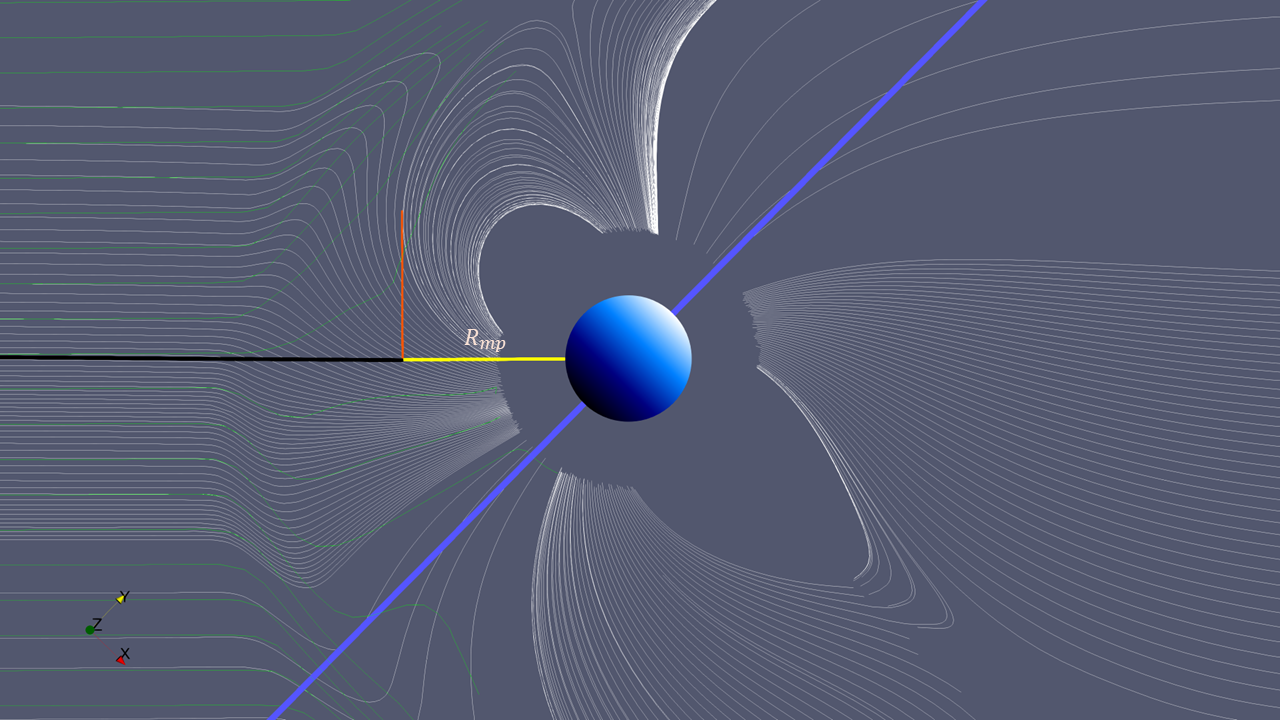}
    \caption{Illustration of how the planetary magnetopause standoff distance \( R_{\mathrm{msd}} \) is measured in our simulations. The distance is defined as the last closed planetary magnetic field line along the star-planet line (here shown in yellow). The black and blue lines represent the radial direction and planetary magnetic axis, respectively.}
    \label{fig:Rmsd_method}
\end{figure}

The procedure adopted to determine the magnetopause standoff distance $R_{\mathrm{msd}}$ in this study is illustrated in Fig.\ref{fig:Rmsd_method}.

\begin{figure*}[htbp]
  \centering
  \begin{subfigure}[b]{0.48\textwidth}
    \includegraphics[width=\textwidth]{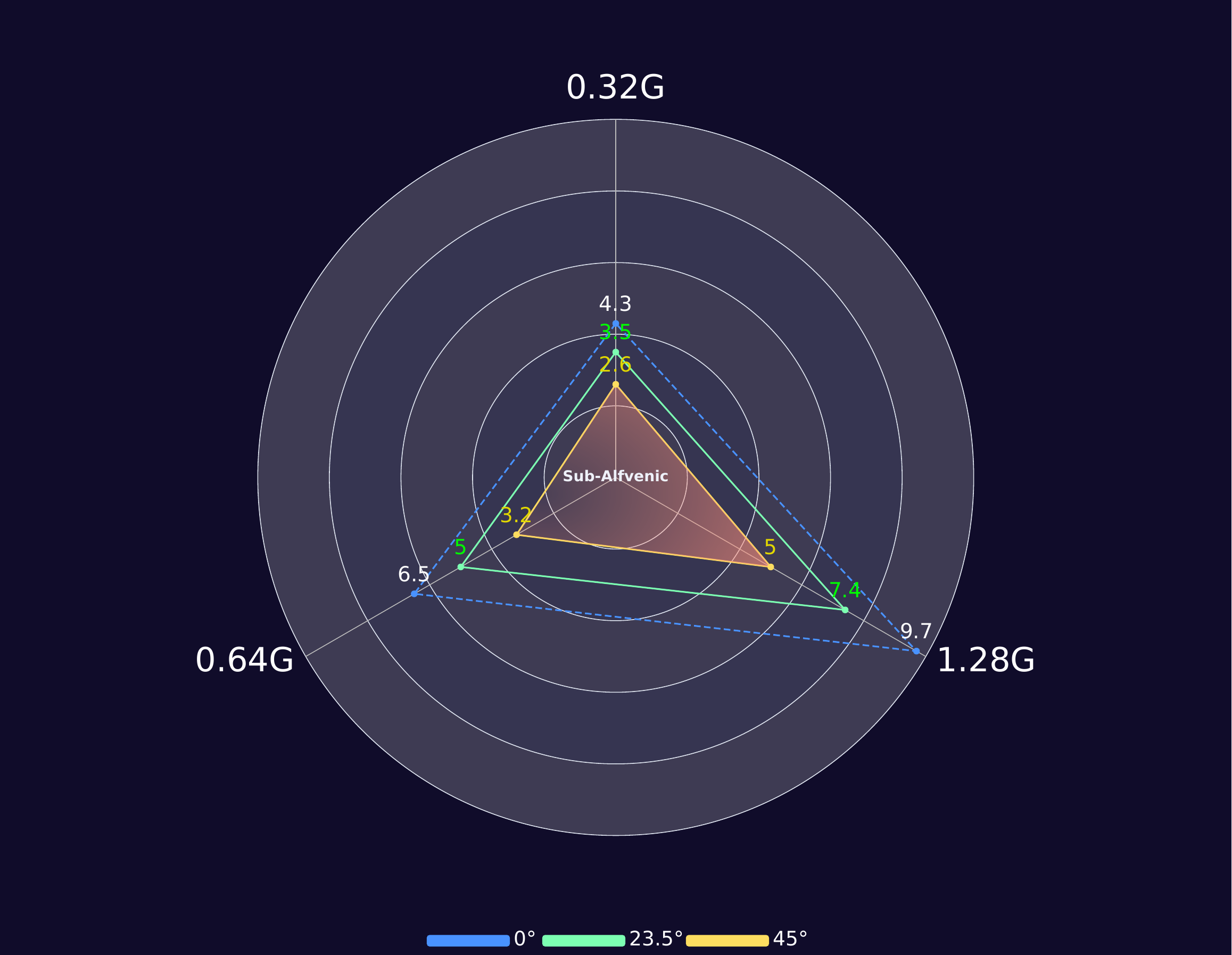}
    \caption{Sub-Alfvénic}
    \label{fig:sub}
  \end{subfigure}
  \hfill
  \begin{subfigure}[b]{0.48\textwidth}
    \includegraphics[width=\textwidth]{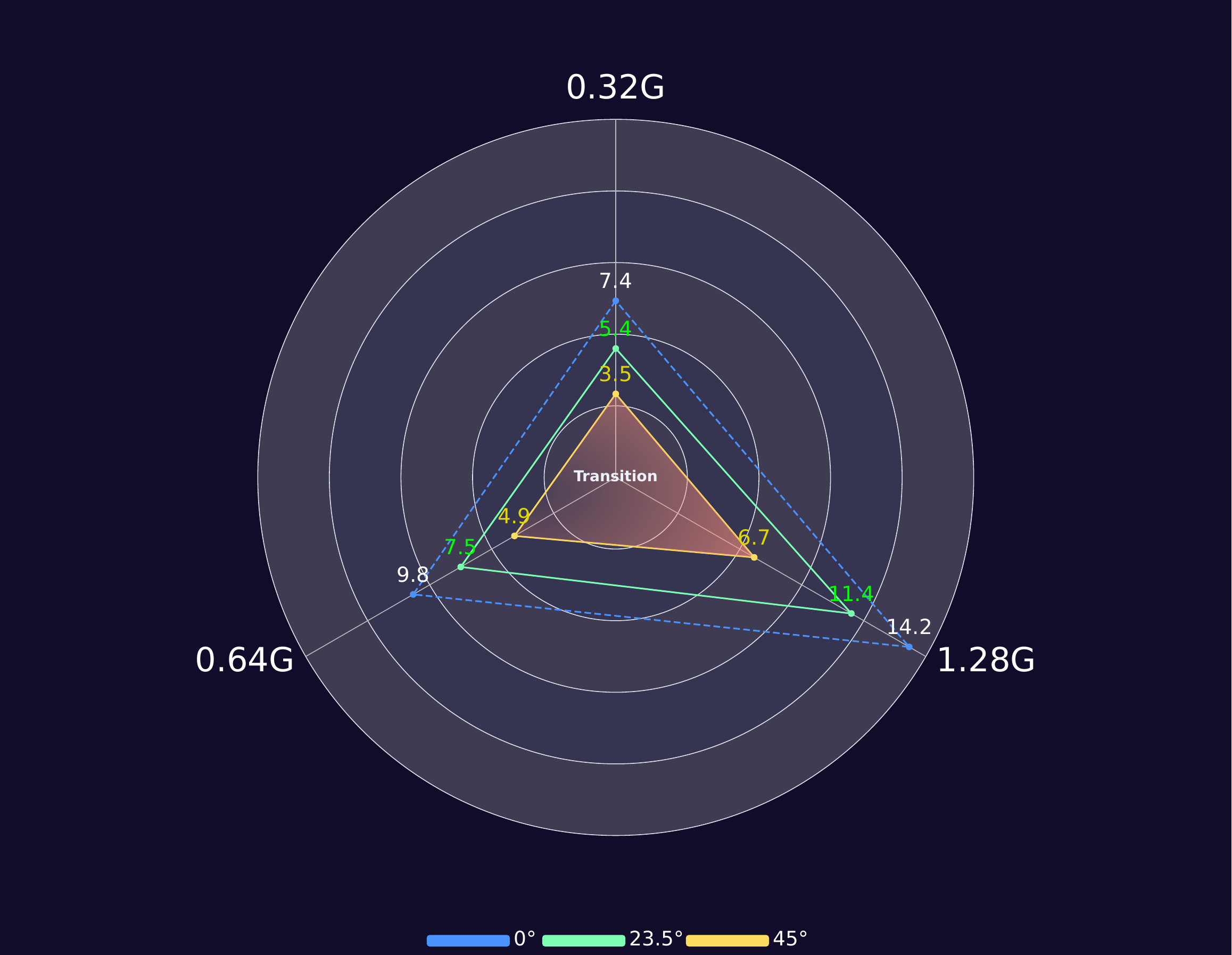}
    \caption{Transition}
    \label{fig:transition}
  \end{subfigure}

  \vspace{0.4cm} % 控制上下间距

  \begin{subfigure}[b]{0.48\textwidth}
    \includegraphics[width=\textwidth]{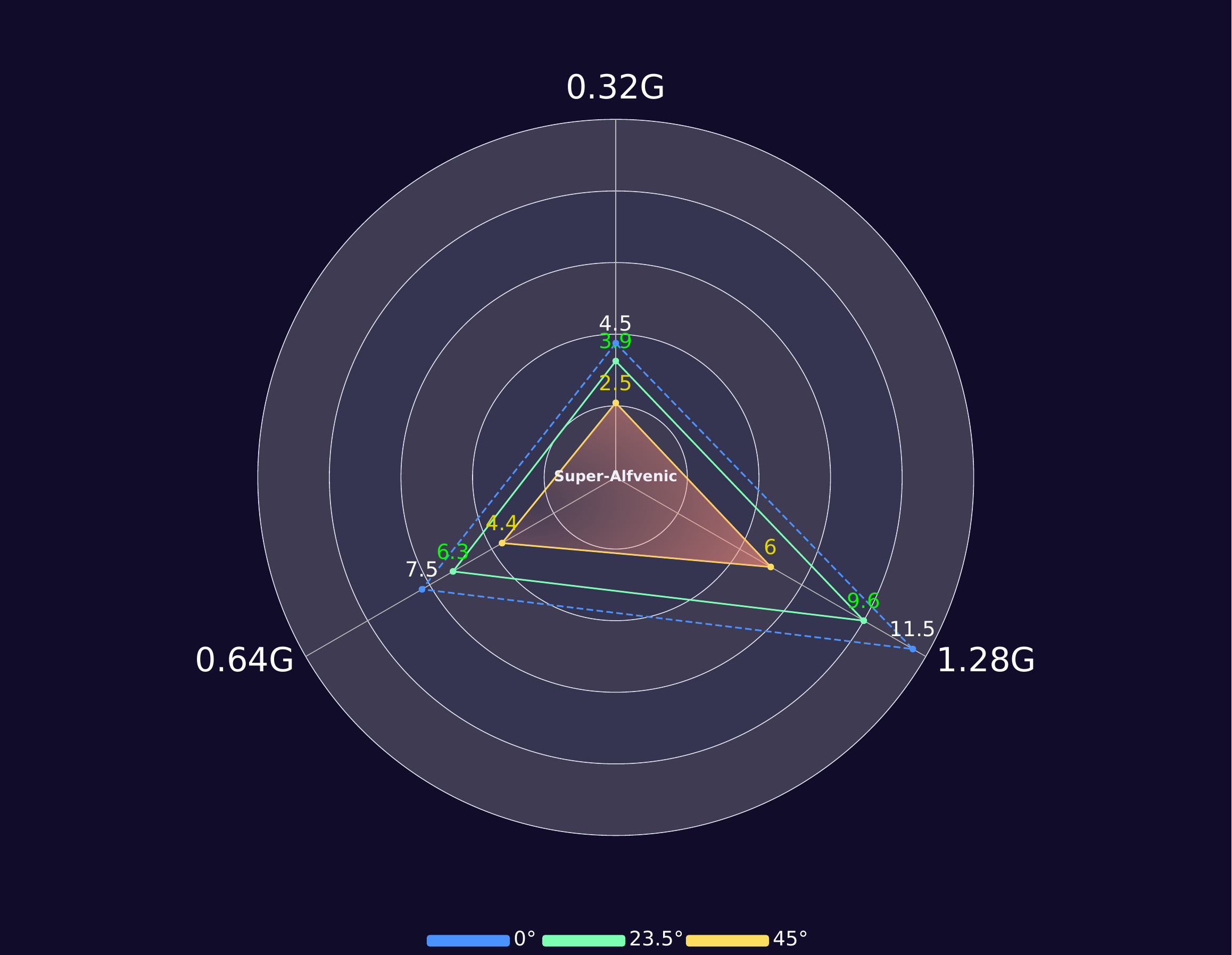}
    \caption{Super-Alfvénic}
    \label{fig:super}
  \end{subfigure}
  \hfill
  \begin{subfigure}[b]{0.48\textwidth}
    \includegraphics[width=\textwidth]{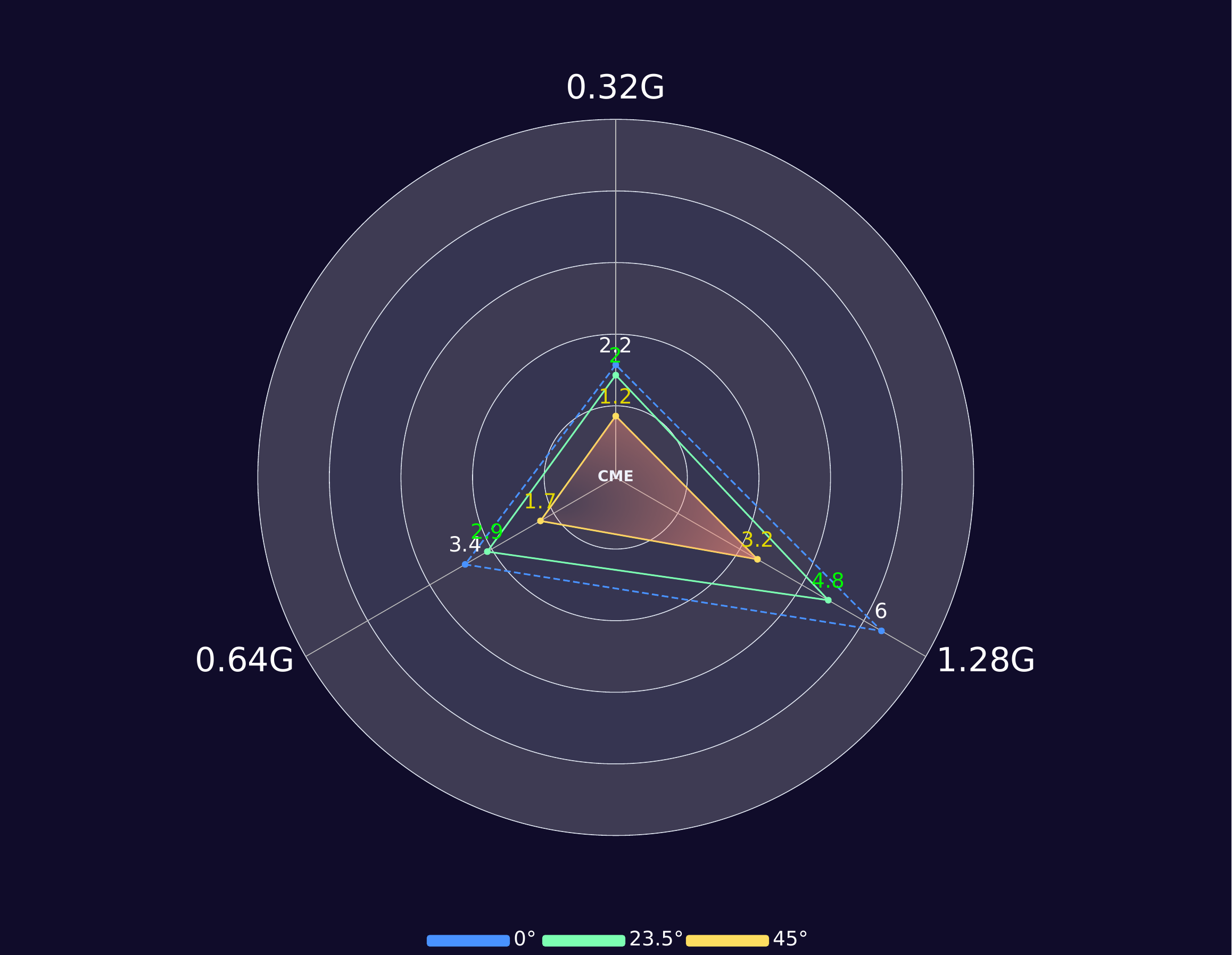}
    \caption{CME-like}
    \label{fig:cme}
  \end{subfigure}
    \caption{Magnetopause standoff distance $R_{\mathrm{msd}}$ of TRAPPIST-1e under four different stellar wind conditions: (a) sub-Alfvénic, (b) transition, (c) super-Alfvénic, and (d) CME-like scenarios. Each polar diagram shows $R_{\mathrm{msd}}$ as a function of magnetic dipole strength ($B_p$ = 0.32, 0.64, 1.28 G) and tilt angle ($0^\circ$, $23.5^\circ$, $45^\circ$). Radial distances are in planetary radii $R_e$. In all scenarios, the magnetopause distance increases with $B_p$ and decreases with magnetic axis tilt. The transition case exhibits the largest magnetosphere extension, while the CME scenario shows extreme compression, with $R_{\mathrm{msd}}$ dropping below 1.5~$R_e$ in the most unfavorable configurations, indicating severe habitability constraints. These results demonstrate the critical interplay between planetary magnetic geometry and external space weather forcing in shaping magnetospheric protection.}
    \label{fig:rmp_radar}
\end{figure*}

The simulation results for the magnetopause standoff distance, $R_{\mathrm{msd}}$, are illustrated in Figure\ref{fig:rmp_radar}, which shows the outcomes under four different space weather scenarios, each evaluated with varying magnetic field strengths and tilt angles. For each condition, the simulations are organized to separately assess the effects of magnetic tilt and field strength on planetary habitability: first by fixing the planetary magnetic field and varying the tilt, and then by fixing the tilt and varying the magnetic field. These dual dependencies are visualized in radar plots, where both parameters can be interpreted simultaneously.

As seen in all four regimes, $R_{\mathrm{msd}}$ decreases monotonically with increasing tilt angle and increases with stronger planetary magnetic fields. In favorable cases, the standoff distance can exceed 10~$R_e$, which is sufficient to shield the planet from harmful stellar wind impacts, indicating a magnetospheric configuration supportive of habitability. The radial distance in each radar plot indicates the magnitude of $R_{\mathrm{msd}}$, and the three vertices correspond to three magnetic field strengths (0.32G, 0.64G, and 1.28~G) and three tilt angles ($0^\circ$, $23.5^\circ$, and $45^\circ$).

In the Sub-Alfvénic case (top left), where the stellar magnetic pressure dominates (with $B_{\star} = 1200$G), the standoff distance reaches up to 9.7$R_e$, while under a weak field of 0.32G and a tilt of $45^\circ$, it drops to 2.6$R_e$, approaching critical thresholds. This demonstrates that even in a magnetically dominated wind environment, weak planetary fields and large tilts can significantly compromise atmospheric protection.

In the Transition regime (top right), where the stellar magnetic field is halved to 600G, the magnetosphere exhibits the widest expansion among all cases. With a 1.28G planetary field and zero tilt, $R_{\mathrm{msd}}$ can reach 14.2~$R_e$. However, increasing the tilt systematically reduces the standoff distance, reinforcing the trend that oblique fields weaken the planet’s magnetic shielding. Stronger magnetic fields, on the other hand, robustly enhance magnetospheric protection.

Under the Super-Alfvénic condition (bottom left), where the stellar wind dynamic pressure exceeds the magnetic pressure, the magnetosphere becomes moderately compressed. The maximum $R_{\mathrm{msd}}$ reaches 11.5~$R_e$ for a field strength of 1.28~G and $0^\circ$ tilt. This emphasizes the critical role of ram pressure in shaping the magnetospheric boundary.

The most extreme scenario (bottom right) corresponds to CME-like conditions, as outlined in Table~\ref{table1}, with greatly increased solar wind density, velocity, and IMF intensity. Here, the magnetosphere is severely compressed: the maximum $R_{\mathrm{msd}}$ only reaches about 6~$R_e$, and in the worst case (0.32G, $45^\circ$ tilt), it falls below the model’s inner boundary limit of 1.5$R_e$, reaching just 1.2~$R_e$. This implies the planet would be directly exposed to the stellar wind, effectively losing magnetic protection even under moderate field strengths, particularly at high tilts.

We note that our simulation results, although broadly consistent with the theoretical trend of $R_{\mathrm{msd}} \propto B_p^{1/3}$, deviate significantly in magnitude. While the standoff distance increases monotonically with field strength, the rate of increase does not strictly follow the one-third power law. This discrepancy arises from the assumptions embedded in the theoretical model, which idealizes the magnetosphere as symmetric, static, and purely dipolar. It ignores key influences such as reconnection topology, magnetic field orientation, and three-dimensional structure.

In our simulations, the IMF is oriented northward to emulate the planetary magnetic tilt. This affects both the reconnection efficiency and the global magnetospheric topology, leading to deviations from the static pressure balance. These effects are physically meaningful: the evolution of planetary magnetospheres is shaped by reconnection, magnetosheath dynamics, and boundary instabilities, all of which are naturally included in full 3D MHD models. Consequently, while the $B_p^{1/3}$ law provides a useful estimate, it has limitations in dynamically evolving or extreme space weather environments.

%---------------------------------------------------------------------
\section{Radio Emission from TRAPPIST-1e’s Magnetosphere under Various Space Weather Conditions}
\label{section4}

The detection of radio emission generated by the magnetospheres of exoplanets is likely the only currently viable observational method for directly constraining their intrinsic magnetic fields. This has profound implications for understanding the internal structure and dynamics of exoplanets. Once the magnetic properties of an exoplanet are inferred, the space weather environment imposed by its host star along the planet’s orbit can be more thoroughly characterized. As a result, the strength of radio emission from the exoplanet’s magnetosphere is closely coupled with the ambient stellar wind conditions. Furthermore, numerical modeling of radio emission provides theoretical support for future observational campaigns targeting exoplanetary systems—particularly Earth-like planets or hot Jupiters—via radio interferometry, where magnetospheric star–planet interaction (SPI) may be detectable.

In this section, we estimate the radio power emitted from the dayside magnetopause reconnection regions of TRAPPIST-1e under four different space weather conditions. The radio emission is computed based on our MHD simulations of the interaction between the stellar wind of the M-dwarf TRAPPIST-1 and the planetary magnetic field of TRAPPIST-1e. This analysis represents one of the core goals of this study.

We adopt the radio-magnetic Bode’s law, in which the incident magnetized energy flux and the strength of the planetary magnetic obstacle determine the emitted radio power, expressed as $P_{\mathrm{R}} = \beta P_{\mathrm{B}}$, where $\beta$ is the empirical energy conversion efficiency, typically ranging from $(2$–$10) \times 10^{-3}$ \citep{Zarka2018}. The dissipated power $P_{\mathrm{B}}$ corresponds to the Poynting flux divergence in the dayside magnetosphere, and is evaluated numerically by integrating over the volume between the bow shock nose and the magnetopause:
\begin{equation}
    P_R=\beta P_B=\beta \int_V \nabla\cdot\frac{(V\times B)\times B}{4\pi} dV
\end{equation}
This integration method follows the approach described in \citet{Luis2024}, where $P_{\mathrm{B}}$ is computed directly from the simulation outputs, and the only empirical assumption is the value of $\beta$.

It is worth noting that cyclotron maser emission is produced by electrons accelerated along planetary magnetic field lines. This emission can originate either near the planet (planetary radio) or propagate along Alfvén wings toward the star (stellar SPI radio). The latter typically occurs only in the sub-Alfvénic regime, where one of the Alfvén wings can effectively transport momentum and energy to the star. In contrast, planetary radio emission can be generated under both sub- and super-Alfvénic conditions. Our focus here is on the planetary radio emission arising from the magnetopause reconnection regions, especially the dominant Poynting flux from the dayside.

\begin{figure}
    \centering
    \includegraphics[width=\linewidth]{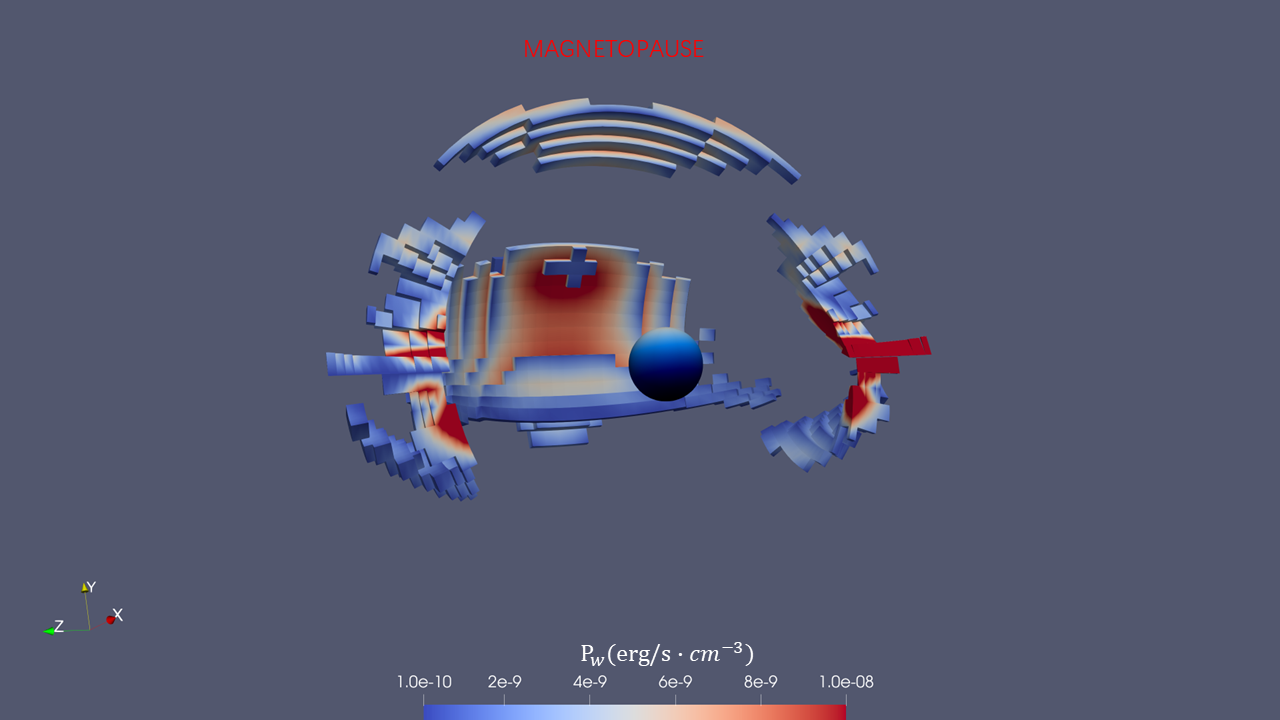}\
    \includegraphics[width=\linewidth]{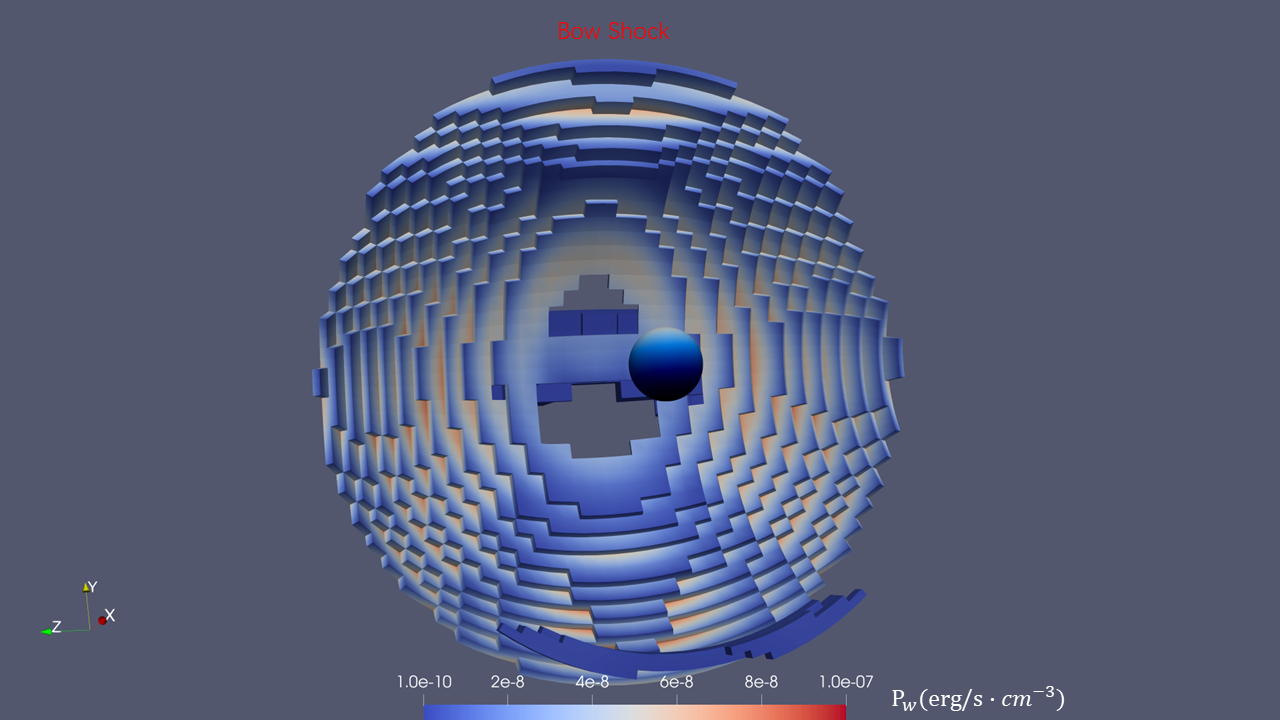}
    \caption{Contribution to the radio emission from the Magnetopause (upper) and BowShock (bottom) region of TRAPPIST-1e derived from our simulation.The volumetric distributions represent the dissipated magnetic energy flux (Poynting flux, $P_w$), highlighting the intensity and spatial structure of energy dissipation responsible for planetary radio emission under stellar wind interaction conditions }
    \label{fig:radio_emission}
\end{figure}

To avoid artificial artifacts near the numerical boundary, we define the integration region with care, excluding volumes too close to the inner boundary. For simulations with a bow shock, the domain encloses all high-density post-shock regions formed due to the collision between the stellar wind and planetary field. In Figure~\ref{fig:radio_emission}, we illustrate the total radio emission for each case, combining contributions from the magnetopause and bow shock where both are present (e.g., in the super-Alfvénic and CME-like cases).

\begin{figure*}[htbp]
  \centering
  \begin{subfigure}[b]{0.48\textwidth}
    \includegraphics[width=\textwidth]{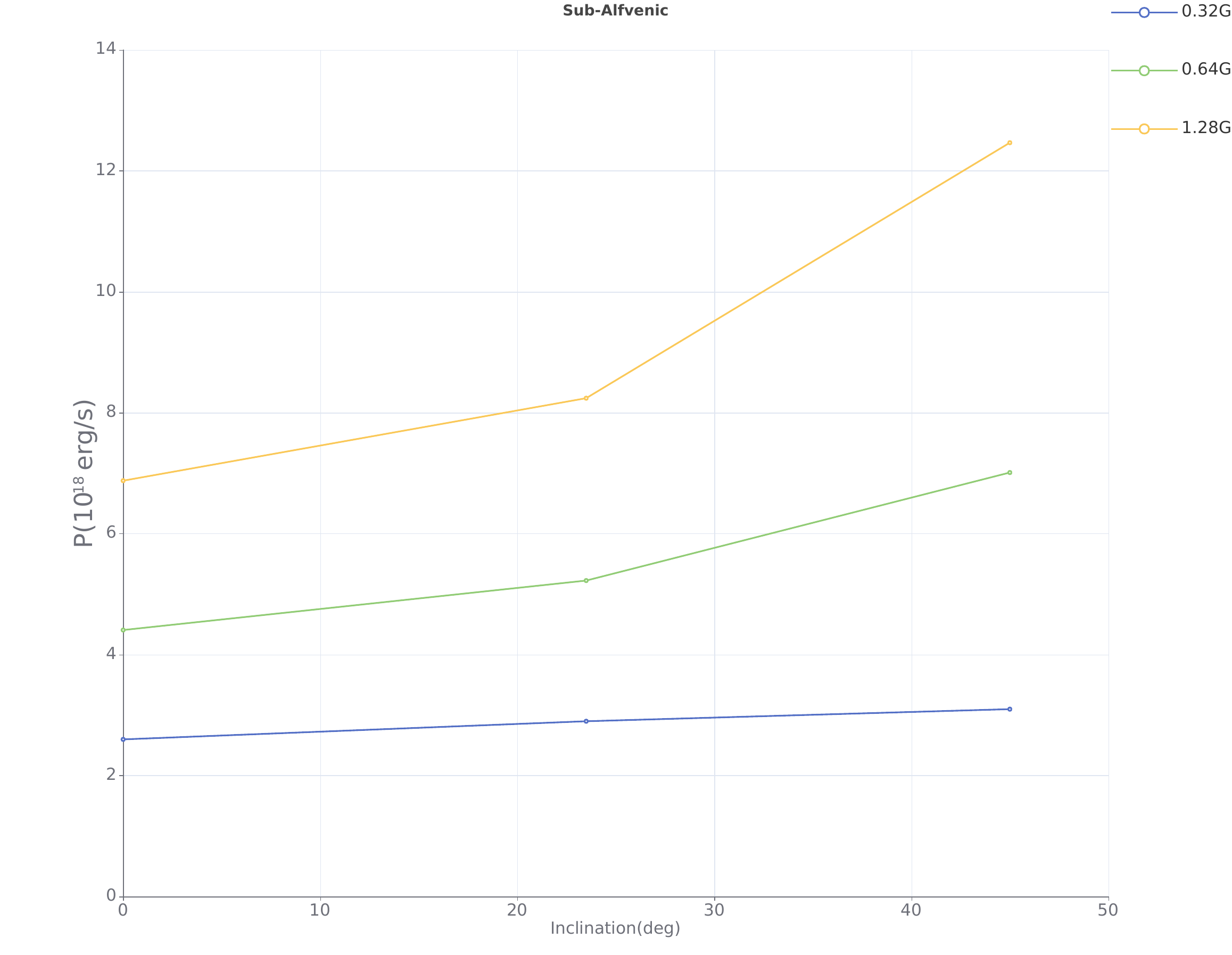}
    \caption{Sub-Alfvénic}
    \label{fig:sub}
  \end{subfigure}
  \hfill
  \begin{subfigure}[b]{0.48\textwidth}
    \includegraphics[width=\textwidth]{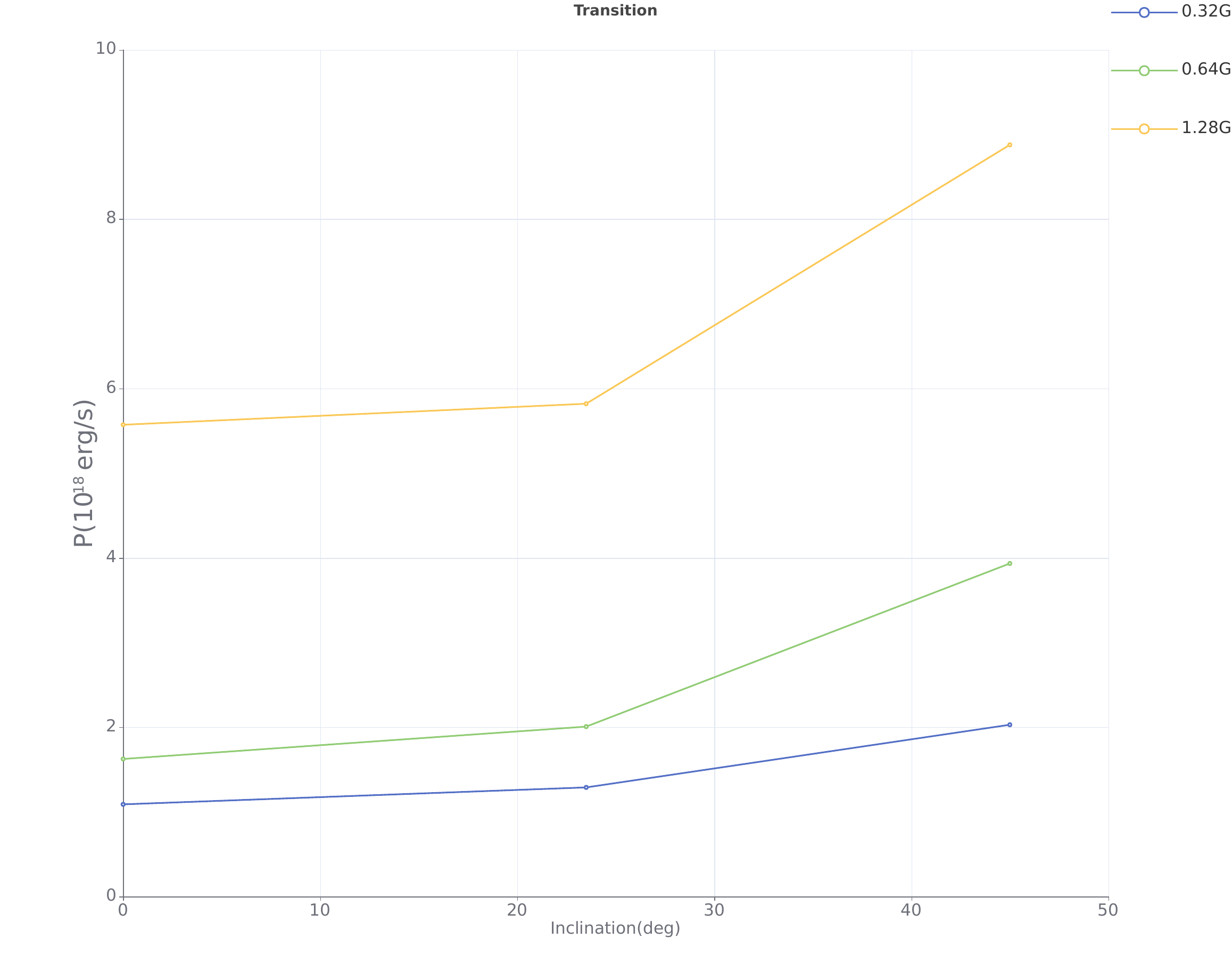}
    \caption{Transition}
    \label{fig:transition}
  \end{subfigure}

  \vspace{0.4cm} % 控制上下间距

  \begin{subfigure}[b]{0.48\textwidth}
    \includegraphics[width=\textwidth]{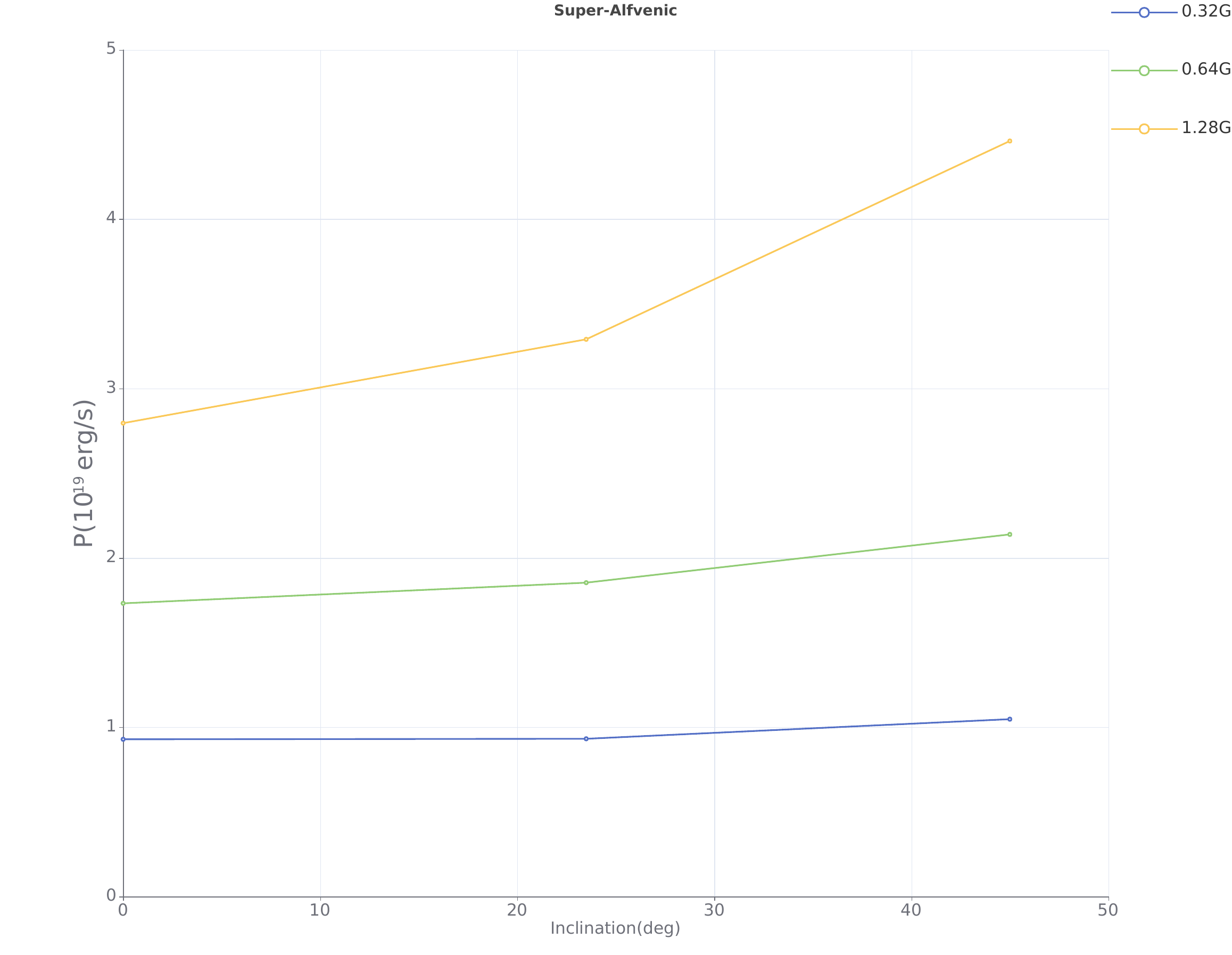}
    \caption{Super-Alfvénic}
    \label{fig:super}
  \end{subfigure}
  \hfill
  \begin{subfigure}[b]{0.48\textwidth}
    \includegraphics[width=\textwidth]{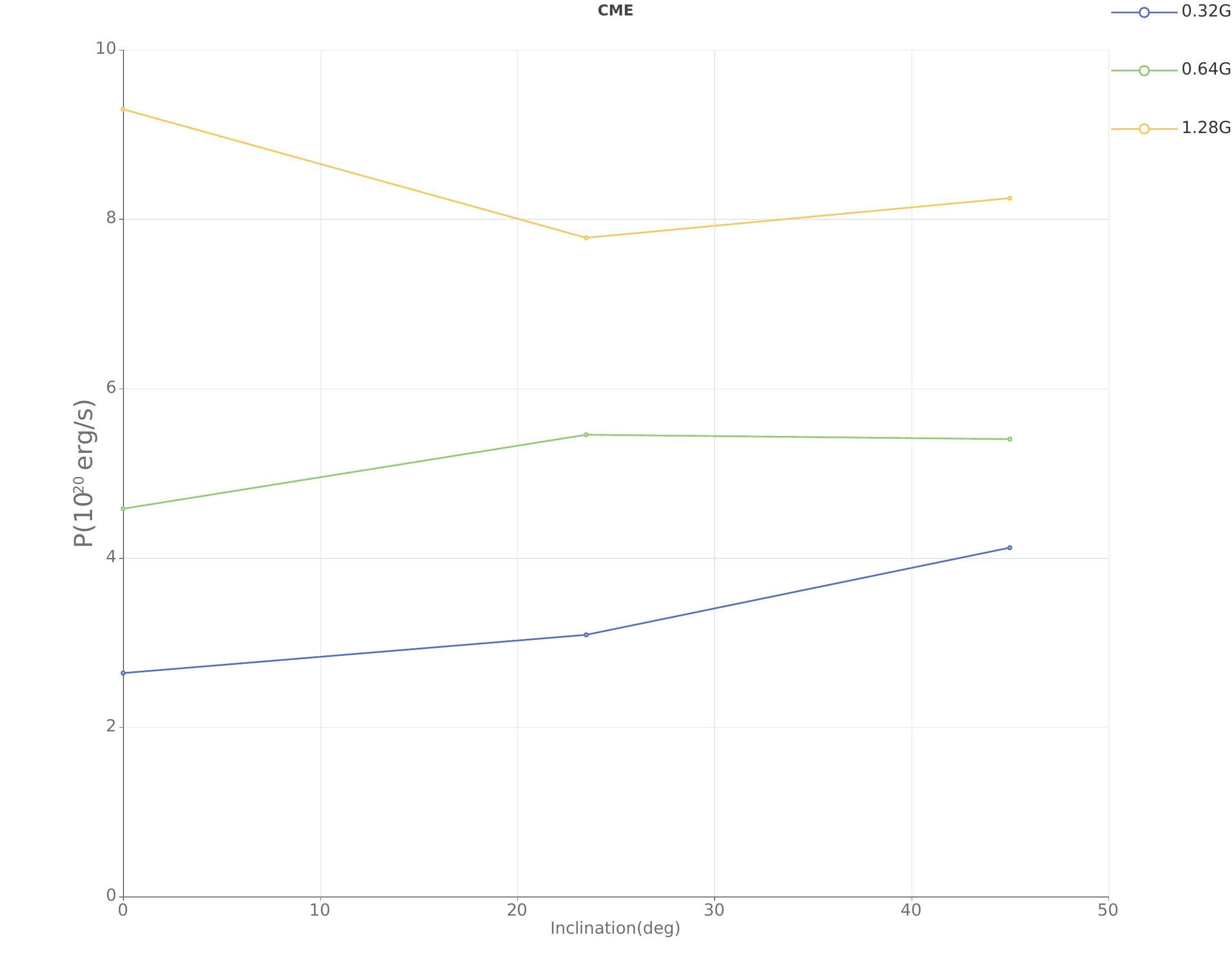}
    \caption{CME-like}
    \label{fig:cme}
  \end{subfigure}
    \caption{ Radio emission plots produced by the planets Sub-Alfvénic, Transition, Super-Alfvénic , CME-like, and TRAPPIST-1e under four different space weather conditions, as a function of inclination and magnetic field strength, with the horizontal axis being the inclination of the magnetic axis, the different colors representing different magnetic field strengths, and the vertical axis being the radiated power ($erg/s$), and the results were obtained using the empirical factor $\beta$.}
    \label{fig:radio}
\end{figure*}

Figure~\ref{fig:radio} presents the simulated radio power $P_{\mathrm{R}}$ of TRAPPIST-1e under four different space weather regimes: sub-Alfvénic, transition, super-Alfvénic, and CME-like. Each panel shows how $P_{\mathrm{R}}$ varies with planetary magnetic field strength and tilt angle.

In the sub-Alfvénic and transition regimes, no bow shock forms. Radio emission therefore arises solely from reconnection at the planetary magnetopause. In the super-Alfvénic regime, both a magnetopause and bow shock exist, and energetic electrons are generated in both regions. This results in combined radio contributions linked to IMF pile-up and bending. For simplicity, we report only the total emission (magnetopause plus bow shock) in all panels of Figure\ref{fig:radio}.

%------------------------------------------------------------------

%-----------------------------------------------------------------
In the sub-Alfvénic case (Figure~\ref{fig:sub}), the maximum radio power reaches $1.25 \times 10^{19}$ erg s$^{-1}$ for $B_p = 1.28$ G and a tilt angle of $45^\circ$. The emission decreases with lower $B_p$ and smaller tilt angles, indicating the crucial role of magnetic configuration in determining emission efficiency. In the transition case ($M_A \sim 1$; Figure~\ref{fig:transition}), the maximum radio output is the lowest among the calm space weather cases, reaching only $8.9 \times 10^{18}$ erg s$^{-1}$ under strong field and large tilt. This reduction may be attributed to weaker stellar wind forcing and topological distortion near the magnetopause, reducing reconnection efficiency.

In the super-Alfvénic regime (Figure~\ref{fig:super}), enhanced dynamic pressure produces a more compact and active magnetosphere. The maximum emission reaches $4.5 \times 10^{19}$ erg s$^{-1}$ for $B_p = 1.28$ G, while for weaker fields, the emission drops below $1.0 \times 10^{19}$ erg s$^{-1}$. The emission increases with tilt, implying stronger reconnection along displaced magnetic shear regions.

In the CME-like regime (Figure~\ref{fig:cme}), the stellar wind exhibits significantly higher speed, density, and IMF strength. The radio emission peaks at $9 \times 10^{20}$ erg s$^{-1}$—the strongest among all scenarios—reflecting extreme compression and reconnection enhancement driven by CME conditions.

In summary, our simulations demonstrate that TRAPPIST-1e’s radio emission is highly sensitive to both internal planetary parameters (magnetic field strength and tilt) and external space weather drivers. The CME-like scenario in particular stands out as a potential high-emission target for exoplanetary radio detection.

\section{Conclusions}
\label{section5}

In this study, we have conducted a comprehensive investigation of the magnetospheric response and radio emission properties of the Earth-sized exoplanet TRAPPIST-1e under a range of stellar wind conditions, using three-dimensional magnetohydrodynamic (MHD) simulations implemented with the PLUTO 4.4 code. Our simulation framework captures both the magnetic and fluid dynamics of the planetary magnetosphere in interaction with the interplanetary magnetic field (IMF), providing insights into the standoff distance of the magnetopause and the associated Poynting flux that drives radio emission.

We evaluated four representative space weather scenarios: the sub-Alfvénic regime, a transitional regime near the Alfvén surface, a super-Alfvénic regime, and an extreme CME-like event. In each case, we varied planetary magnetic field strengths and magnetic axis tilt angles to explore how internal planetary configuration modulates the resulting magnetospheric structure and energy dissipation.

Our simulations reveal that the magnetopause standoff distance $R_{\mathrm{msd}}$ grows monotonically with increasing planetary magnetic field strength, though the rate of this growth does not strictly follow the canonical $R_{\mathrm{msd}} \propto B_p^{1/3}$ law predicted by idealized pressure balance models. This deviation arises due to complex 3D magnetospheric topology, asymmetric reconnection, and the presence of boundary instabilities. These effects  indicating that realistic modeling of exoplanetary space environments must go beyond static pressure balance to include topological and dynamic phenomena.

A key component of this work is the estimation of radio emission arising from dayside magnetospheric reconnection, using the radio-magnetic Bode’s law: $P_{\mathrm{R}} = \beta P_{\mathrm{B}}$, with $\beta = 2 \times 10^{-3}$ as the adopted efficiency. We integrated the divergence of the Poynting vector within the region bounded by the bow shock and magnetopause to determine the dissipated magnetic energy available for particle acceleration. These estimates reveal that TRAPPIST-1e could generate high levels of coherent radio emission under suitable stellar wind conditions. For example, under extreme CME-like conditions, the total radio power reaches as high as $9 \times 10^{20}$ erg/s, while in the more moderate sub-Alfvénic case, peak emission is around $1.2 \times 10^{19}$ erg/s for favorable planetary parameters and can reach $4.5\times10^{19}$erg/s for super-Alfv\'enic.

We find that planetary magnetic field strength exerts a primary control on radio power output, but that tilt angle also plays a significant modulating role, especially in sub- and super-Alfvénic regimes.Despite the promising levels of radio emission predicted by our models, the majority of these signals are expected to fall below the ionospheric cutoff frequency of Earth (typically around 10 MHz). This implies that such emissions would not be detectable by ground-based radio telescopes if TRAPPIST-1e possesses a magnetic field strength similar to Earth's ($\sim 0.3$ G). However, space-based or lunar-surface radio arrays operating at low frequencies could overcome this limitation and represent a compelling avenue for future detection efforts. Our results also underscore the potential detectability of more magnetically active exoplanets, such as hot Jupiters, which may produce emission above the ionospheric cutoff.

Looking forward, space missions like China's proposed ``Discovering the Sky at the Longest Wavelength'' (DSL) array on the lunar far side provide promising platforms for such low-frequency exoplanetary radio studies \citep{Chen2023DSL}. In parallel, new advances in domestic instrumentation such as the FAST Core Array project \citep{Jiang2024FAST} could enhance detection capability in higher-frequency regimes, aiding the search for bursty or harmonically upshifted exoplanetary emissions.In summary, our study demonstrates that the interaction between TRAPPIST-1e and the stellar wind environment of TRAPPIST-1 has the potential to generate significant magnetospheric radio signals, depending on planetary magnetic parameters and external space weather forcing. These signals, though likely inaccessible to ground-based observatories, offer exciting prospects for future space-based detection and contribute to our broader understanding of planetary magnetism, habitability, and star–planet interactions.

\begin{acknowledgements}
      This work is supported by National Natural Science Foundation of China ( Grant No.42274221), National Key R\&D Program of China (No. 2022YFF0503800), and Center for Computational Science and Engineering at Southern University of Science and Technology.
\end{acknowledgements}

% WARNING
%-------------------------------------------------------------------
% Please note that we have included the references to the file aa.dem in
% order to compile it, but we ask you to:
%
% - use BibTeX with the regular commands:
%   \bibliographystyle{aa} % style aa.bst
%   \bibliography{Yourfile} % your references Yourfile.bib
%
% - join the .bib files when you upload your source files
%-------------------------------------------------------------------

\end{document}